\documentclass[journal, onecolumn]{IEEEtran}

\usepackage[numbers]{natbib}

\usepackage{enumitem}
\usepackage{subcaption}
\usepackage{algorithm}
\usepackage{algpseudocode}
\usepackage{multirow}
\usepackage{amsmath}
\usepackage{xcolor}
\usepackage{soul}
\usepackage{pbox}
\usepackage{lipsum}
\usepackage{booktabs}
\usepackage[final]{changes}
\definechangesauthor[name=Reviewer1-1, color=blue]{U}
\definechangesauthor[name=Reviewer1-2, color=green]{V}
\definechangesauthor[name=Reviewer2-1, color=red]{D}
\definechangesauthor[name=Reviewer2-2, color=orange]{E}

\begin{document}



\title{Resource Allocation for Dataflow Applications in FANETs using Anypath Routing}  



%

\author{Juan José López Escobar, Manuel Ricardo, Rui Campos, Felipe Gil-Castiñeira, Rebeca P. Díaz-Redondo
\thanks{Juan José López Escobar (juanjo@det.uvigo.es), Felipe Gil-Castiñeira (xil@gti.uvigo.es) and Rebeca P. Díaz-Redondo (rebeca@det.uvigo.es) are with atlanTTic, Universidade de Vigo; Vigo, 36310, Spain.}
\thanks{Manuel Ricardo (manuel.ricardo@inesctec.pt) and Rui Campos (rui.l.campos@inesctec.pt) are with Institute for Systems and Computer Engineering, Technology and Science (INESC TEC) and Faculdade de Engenharia. Universidade de Porto, Porto, 4200-465, Portugal}}

\maketitle

\begin{abstract}
Management of network resources in advanced IoT applications is a challenging topic due to their distributed nature from the Edge to the Cloud, and the heavy demand of real-time data from many sources to take action in the deployment. FANETs (Flying Ad-hoc Networks) are a clear example of heterogeneous multi-modal use cases, which require strict quality in the network communications, as well as the coordination of the computing capabilities, in order to operate correctly the final service. In this paper, we present a Virtual Network Embedding (VNE) framework designed for the allocation of dataflow applications, composed of nano-services that produce or consume data, in a wireless infrastructure, such as an airborne network. To address the problem, an anypath-based heuristic algorithm that considers the quality demand of the communication between nano-services is proposed, coined as Quality-Revenue Paired Anypath Dataflow VNE (QRPAD-VNE). We also provide a simulation environment for the evaluation of its performance according to the virtual network (VN) request load in the system. Finally, we show the suitability of a multi-parameter framework in conjunction with anypath routing in order to have better performance results that guarantee minimum quality in the wireless communications.
\end{abstract}


\maketitle









\section{Introduction}\label{introduction}
Unmanned aerial vehicles (UAVs) have gone from being restricted to particular scenarios --- due to public safety --- to having a regulated legal framework that allows them to be used in many different domains, both academic and commercial \cite{uav_regulations}, expanding their action areas to military, marine and surveillance purposes.

In this regard, coordinated and simultaneous flying of several drones has emerged to cope with use cases where large geographic areas must be covered to offer a service, known as “flying ad hoc networks” (FANETs) \cite{fanets_survey}. These multi-UAV systems are characterized by mobility and topology changes since (i) devices are moving to provide an adapted service and (ii) they are powered by batteries, which implies an effort to maintain network reliability.

Of special note are emergency use cases where a set of small UAVs collaborate to carry out search and rescue missions, control and manage natural disasters, and take concrete action to help do their job. More precisely, drones can access difficult and hazardous places, having a wider view of the situation and preventing staff members from being involved in an accident, thereby improving success and security. In addition, proper management of natural disasters require the usage of novel IoT sensing and communication technologies for enhancing the prevention, intervention and recovery \cite{survey_wsn_iot_disaster}. Indeed, UAVs usually carry sensors and actuators to interact with the environment, and extract the necessary information to make decisions during the operation, while exploiting the advances in wireless communications to shape new mobile network topologies.

In order to enable these activities through a mesh of UAVs, they must provide enough computational power to collect, store, process and share data. Also, to optimize the performance of the service, the network should adapt intelligently by taking actions according to the dynamic features of the network. The collaborative aspect of the network is important, since it allows combining several sources of information in a distributed application to make the most of all resources of the aerial infrastructure, as it is the case of federated learning in an artificial intelligent (AI) context \cite{federated_learning_uavs}. Besides, UAVs must be able to operate independently of the backbone network either due to the unstable connectivity of FANETs, or the difficulty in satisfying Quality of Service (QoS) requirements such as low latency for truly real-time, high bandwidth demand to prevent core network saturation, or privacy patterns that preclude the use of external cloud services. Thus, in the scenarios addressed by this paper, computing must be flexible and run in the edge flying devices.

\begin{figure}[!ht]
    \centering
    \includegraphics[width=0.6\linewidth]{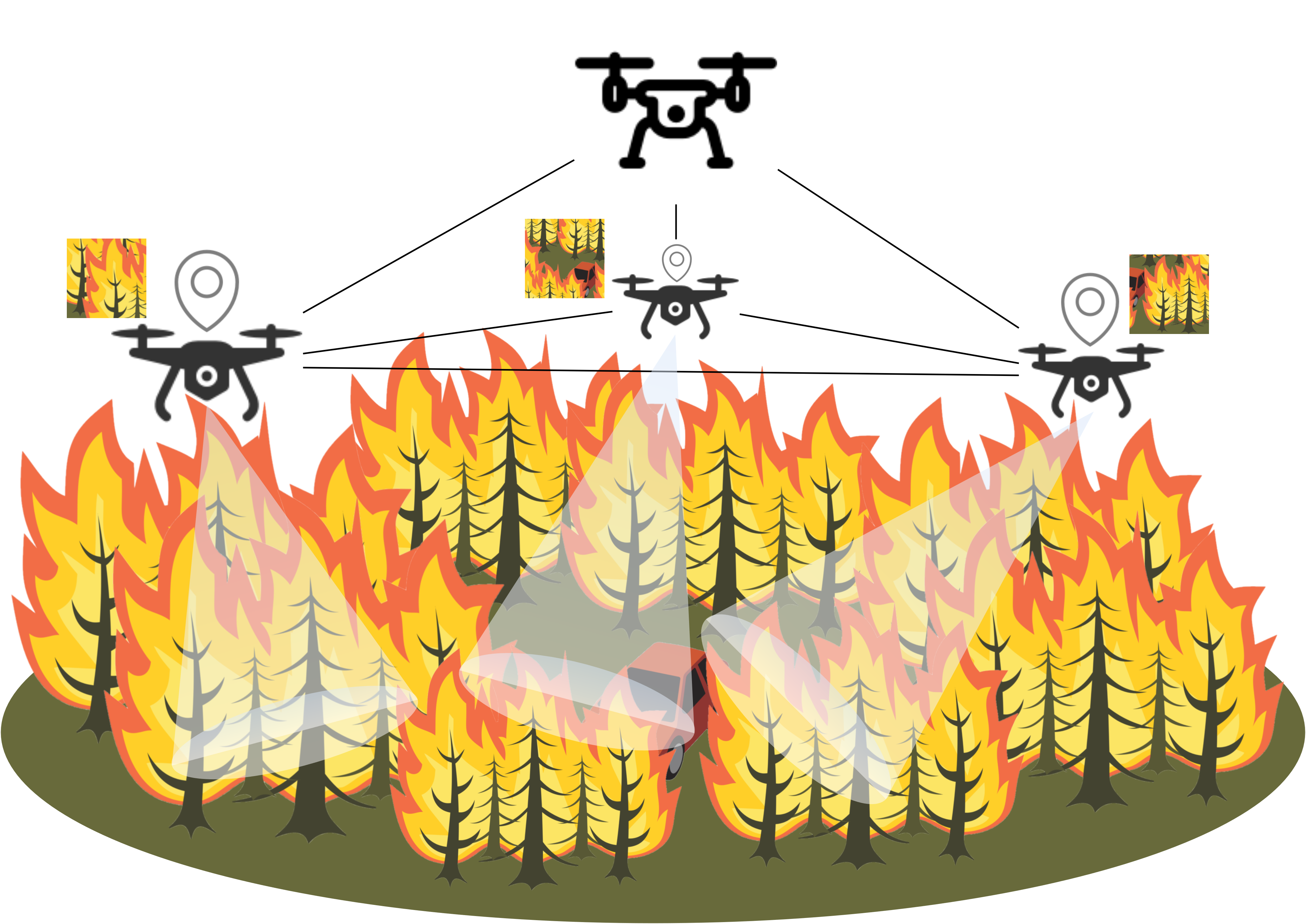}
    \caption{Search and rescue in forest fire using FANETs.}
    \label{fig:fanet_surveillance}
\end{figure}

As depicted in Figure \ref{fig:fanet_surveillance}, an example is multimodal video surveillance in search and rescue operations, such as forest fires, which may be used to identify and locate people from the air, thus making the job of staff members more effective and efficient. In this case, a drone swarm is spread out over different areas of the forest to record a video flow, which is analyzed in real time and shared among the UAV network to adjust the surveillance area and detect the desired target. The results of the video processing may be added to other information sources such as cellular signal or GPS location for higher accuracy \cite{gsm_phone_location, au-air_multimodal_surveillance}.

Any distributed application running on top of the FANET infrastructure may be partitioned among the UAVs considering constraints, such as power consumption or reaction time limit. Processes of an application may be moved from one UAV to another depending on factors such as end-user behavior. For instance, the main routine to follow a terrestrial target must be deployed in a UAV which has sufficient battery energy and can do the tracking correctly.

Virtual Network Embedding (VNE) is a promising methodology to make the most of all available resources and dynamically assign and move specific tasks \cite{vne_survey}. This technique focuses on finding an optimal assignment of the requested set of resources to the available physical resources. In this way, multiple virtual networks are deployed on top of the same physical network according to application computing and communication requirements. Despite being widely studied in recent years for virtualization enhancement in wired networks, VNE has left out wireless and mobile networks, where it may have a great impact on mobile wireless sensor networks to leverage intelligent dynamic services on top.

More specifically, VNE can be easily applied to FANET deployments since they will operate distributed services, which can follow the dataflow programming model \cite{fog_at_edge_dataflow_node}. Under this premise, a virtual application is composed of several nano-services that perform specific tiny operations \cite{nanoservice} and communicate unidirectionally to send messages and continue with the execution. Then, each task is defined by a set of functional and non-functional requirements that must be supported to allocate the virtual service \cite{fogflow, data-flow_zenoh}.

\begin{figure}[!ht]
    \centering
    \includegraphics[width=0.6\linewidth]{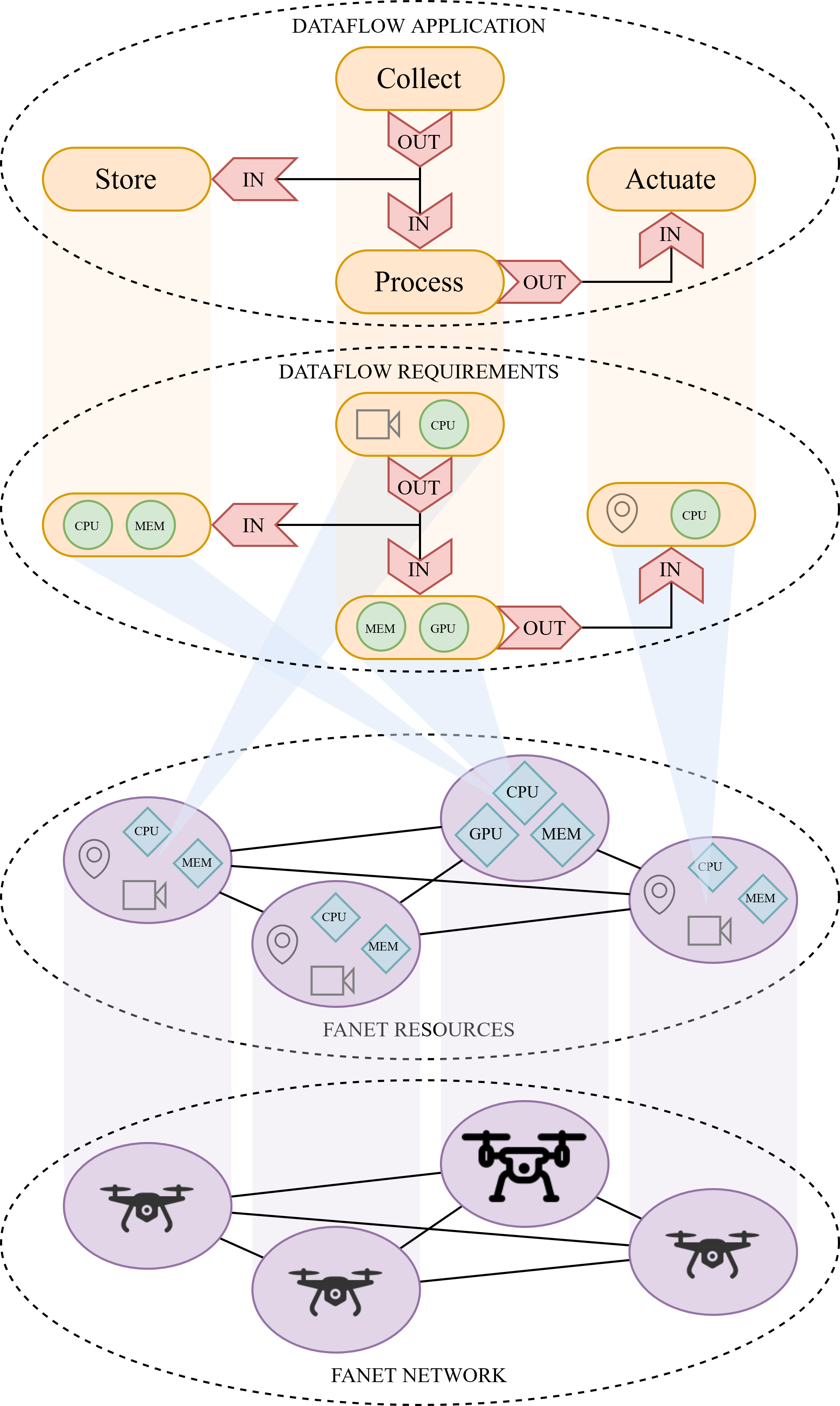}
    \caption{Mapping of a dataflow application to a FANET deployment.}
    \label{fig:dataflow_application_fanet}
\end{figure}

In the particular abovementioned emergency surveillance scenario, the application from Figure  \ref{fig:dataflow_application_fanet} can be mapped to a physical UAV infrastructure, regarding the service composition and requirement description done. The 'Collect' function starts the flow by gathering data from the environment and sending the raw income to the 'Store' function, which saves it. The 'Process' function analyzes the information flow and sends its output to the final 'Actuate' function, which can then enforce a decision at the aerial scenario. In accordance to the amount of resource demand (e.g. CPU, GPU and memory units) and functionalities associated to sensors and actuators, such as video cameras or GPS devices, to execute the specific services, as well as communication specifications, tasks are deployed (as indicated in Figure \ref{fig:dataflow_application_fanet}) using the VNE solution. The 'Store' and 'Process' functions are mapped to the UAV with the best back-haul connection because it can handle powerful computing tasks, while the 'Collect' and 'Actuate' functions have to be deployed in sparse compact UAVs that have sensing capabilities to interact with the environment.

This programming model defines requirements that must be fulfilled to ensure proper operation of each function and adequate quality of communication paths. In addition, it fits perfectly sensing applications, which need to take action constantly with a continuously generated flow of data.

This paper provides a good example of UAV operations in surveillance and rescue works using resource virtualization over FANETs, which is an innovative approach to implement real-time adjustments of the settings according to the needs of the application at each moment. In addition, it introduces a VNE framework that may be applied to different mobile scenarios based on the dataflow programming model, which is suitable for the composition of dynamic and intelligent distributed applications in the advanced Internet of Things world. More specifically, this paper provides three main contributions:

\begin{itemize}
    \item A mathematical model for describing dataflow applications and their resource requirements to be compliant with the VNE methodology, enabling VNE to be used both for final application composition and deployment.
    \item An innovative VNE algorithm for dataflow applications, termed as Quality-Revenue Paired Anypath Dataflow VNE (QRPAD-VNE), that exploits the special features of wireless communication channels in FANETs and tries to optimize the use of resources in order to allocate and execute in the best possible manner.
    \item A reference framework to design and evaluate VNE algorithms for multiple-resource and sensor/actuator-enabled dataflow applications.
\end{itemize}

The paper is structured as follows, in this section the motivation for this research  has been explained and the requirements to solve it have been presented. Section \ref{related_work} presents the existing VNE proposals, which have inspired our solution. Therewith, in Section \ref{model} the problem is modeled using graph theory. In Section \ref{algorithm} the QRPAD-VNE algorithm is presented considering the target scenario, and then it is validated for chosen network settings in Section \ref{evaluation}. Section \ref{conclusion} concludes the work, reflecting possible improvements and future directions.

\section{Related Work}\label{related_work}

Virtualization is regarded as a cutting-edge topic in modern network technologies, such as 5G and IIoT (Industrial Internet of Things), which provides sophisticated techniques that give flexible and straightforward management of resources, as well as exploiting them efficiently at any level of the infrastructure. This is where VNE technology comes in. Indeed, it has been extensively used in service planning for networking functions \cite{vne_algorithms}, and many solutions have been proposed for wired network backbones of the telecom operators \cite{vne_survey}. However, the wireless domain continues to pose an important research challenge, and VNE raises with innovation potential to cope with the special characteristics of wireless multi-hop networks to get the better communication quality \cite{vne_wireless_multihop}, but it also allows the composition of services according to the available resources.

Concretely, the Wireless VNE (WVNE) technology can be modeled as a mathematical problem to be solved by an Integer Linear Programming (ILP) mechanism \cite{multi-hop_wvne}, but it is intractable in large deployments due to its NP-hard nature and appropriate heuristic algorithms are needed to each environment \cite{heuristic_vne}. Otherwise, it implies an operating cost of physical nodes and links that needs to be optimized to yield maximum benefit of virtualization capabilities to allocate resources \cite{benefits_wvne}. Most of WVNE proposals provide convenient node resource modeling and put the spotlight on link peculiarities of wireless broadcasting.

In particular, links are the key difference from traditional wired VNE, since physical phenomena and protocols are applied differently. Wireless links must be characterized in terms of bandwidth capacity, latency or delay, reliability or error rate, transmission power, signal interference, geolocation, etc. Moreover, it is recommended to choose a mapping algorithm that takes advantage of the broadcast and wireless nature to improve performance of the resulting embedding \cite{framework_wireless_vne}.

The study in this field was initiated taking the interference conflict into account to evaluate the suitability of the embedding communication \cite{vne_wireless_multihop, static_wvne}, whether to be aware of topology distribution to deploy services from a central point \cite{topology_awareness_wvne}, or to select the communication links with the less noise according to signal power of neighbors \cite{power_control_multi-cast_wvne}. In addition, in \cite{reconfiguration_channel_wvne} the potential signal collisions are considered to enable communication channel reconfiguration in order to benefit the most of effective available bandwidth. Another relevant aspect relates to the capacity of allocating multiple messages simultaneously in the wireless spectrum to maximize resources and avoid collisions \cite{ofdm_channel_wvne}, as well as the influence of distance between nodes to the quality of the links \cite{spectrum_sharing_wvne}. Overall, wireless links can be formally defined by a metric cost from a combination of their bandwidth, delay and reliability for a simplified view of the embedding problem \cite{topology_qoe_vne, vne_wireless_time_efficient_qos_qoi, multi-service_wvne}.
With respect to the algorithm strategy itself, the mapping from virtual communication links to substrate paths is performed solving a shortest-path graph problem according to the metric cost of the chosen model, either for single or multiple split paths \cite{rethinking_vne_path_splitting}. In \cite{distributed_wvne} a distributed consensus approach is presented for unavailable centrally controlled settings.

Most of these VNE algorithms are solved around a single evaluation and classification of node and link characteristics, but there are several proposals that go one step further and introduce the use of genetic algorithms to overcome the complexity of the corresponding ILP model \cite{ofdm_channel_wvne, link_embedding_ilp_genetic, genetic_wvne}.

Furthermore, a very promising approach that exploits the unique features of wireless mesh networks to improve the efficiency of the embedded communication channel is opportunistic rebroadcasting to have multiple paths for the same data \cite{multicast_service_wvne}. Recently, it has been implemented (a) in industrial environments using anypath routing that considers node locations and packet delivery ratio and bandwidth capacity of the links in order to comply with latency or reliability requirements of WSN applications \cite{application_vne_industrial_wireless}, and also (b) in airborne networks to offer location-aware services, which are deployed following a one-stage algorithm that order the mapping according to interference effect and resource demand \cite{reliable_vne_atn}.

Despite the primary focus of WVNE is link assignment, some research works consider node properties in wireless scenarios. In \cite{multidomain_drones_vne} the selection of nodes is based on the cost of all potential choices in the substrate network, and \cite{survivability_wvne} proposes an algorithm that makes the election according to the cost of executing a task in a node that has a certain degree of failure and the deployed backup entities that may recover it.

Another relevant matter is mobility, since FANET networks are, by nature, potentially mobile. In \cite{dynamic_wvne}, a mechanism to support service mobility through multiple node and path election is explored to migrate services when necessary.

VNE is highly related to current technology trends that will leverage automated and intelligent network management, and it will ease the development of virtualization schemes using reference frameworks \cite{architecture_wvne} for 5G, beyond-5G and advanced IoT environments. On the one hand, it has been integrated with Software-defined networking (SDN) architectures due to the centralized view of the state of the substrate network and running virtual services, such as industrial wireless sensor and IoT networks \cite{intelligent_latency_vne_industrial_wireless, link_reembedding_ilp_genetic_sdn}. On the other hand, it has been investigated for cellular network providers to accomplish QoS conditions, as low latency requirements \cite{low_latency_mobile_vne}, to deploy efficiently MEC instances efficiently \cite{mec_vne}, or to create virtual network operators for mobile users \cite{data-driven_anypath}.

In the light of the above, it is clear that VNE has the potential to revolutionize future intelligence-based networks for IoT and 6G technologies, and particularly anypath VNE strategies have the capability to meet the requirements of intensive wireless sensor applications, and it is envisaged to exploit wireless mesh capabilities.

\section{Network Model}\label{model}

This section aims to model the VNE resource mapping problem of an application in a physical network by taking advantage of the dataflow programming approach.

\begin{figure}[!ht]
    \centering
    \subfloat[Programming model of the dataflow application.]{
        \includegraphics[width=0.6\linewidth]{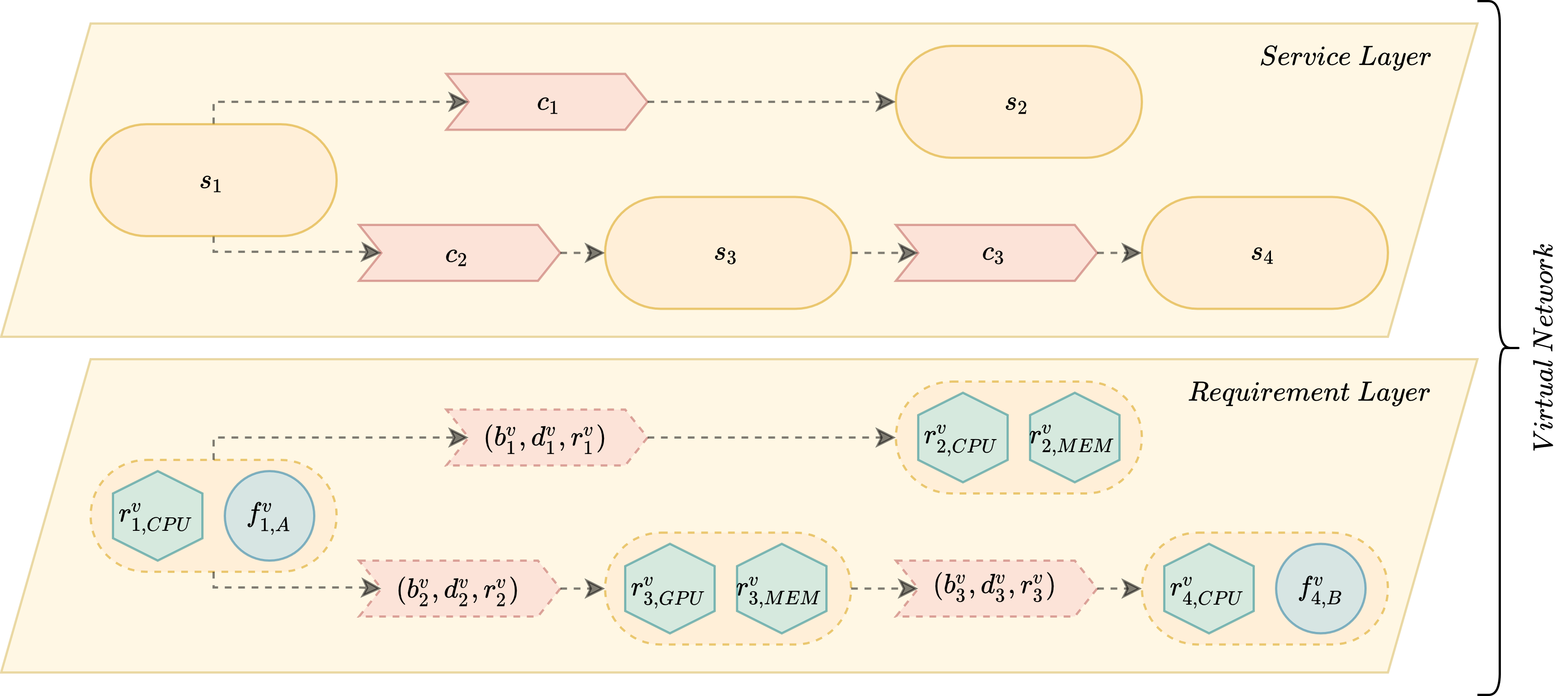}
        \label{fig:application_programming_model}}
        
    \subfloat[Programming model of the physical infrastructure.]{
        \includegraphics[width=0.6\linewidth]{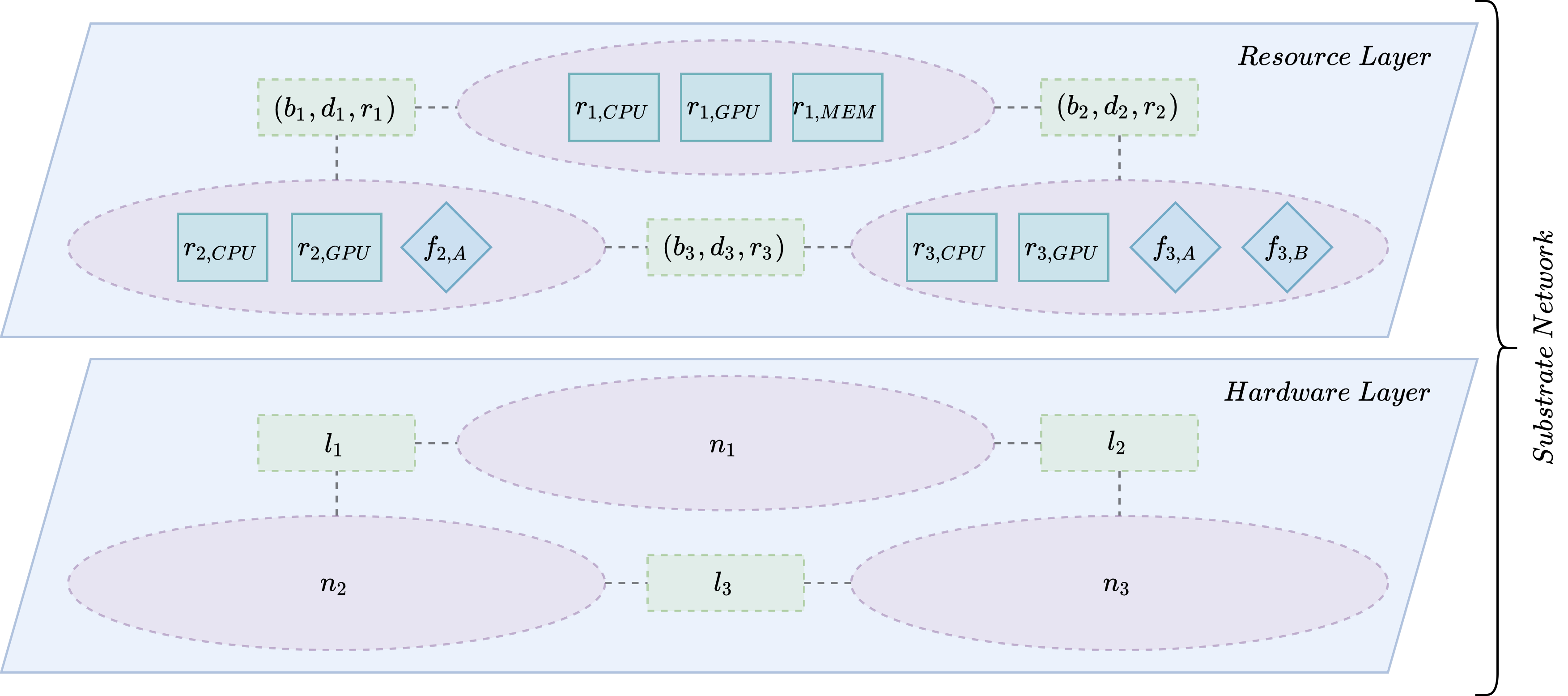}
        \label{fig:physical_programming_model}}
    \caption{Programming model of the dataflow VNE framework.}
    \label{fig:programming_model}
\end{figure}

In order to illustrate the framework designed to target the main problem, we depict the programming model in Figure \ref{fig:programming_model}, which is organized in two parts --- the applications that are requested to be deployed (Figure \ref{fig:application_programming_model}), and the physical infrastructure which offers virtualized resources (Figure \ref{fig:physical_programming_model}). An application can be decomposed into several nano-services that are executed according to the designed programming flow and send the corresponding data flows through the channels (\textit{Service Layer}). All those elements are characterized by the requirements that need to be fulfilled (\textit{Requirement Layer}). In turn, the physical network is formed by the available computing nodes connected between themselves with direct physical links (\textit{Hardware Layer}), and both capabilities from nodes and links are rendered to be used in a shared scenario (\textit{Resource Layer}).

\subsection{Substrate network}

The physical infrastructure is composed of computing nodes connected by direct networked links is denoted by the substrate graph $G^s=(N,L)$, where $N$ and $L$ are the set of nodes and connection links, respectively.

Each substrate node $n \in N$ is defined by a set of attributes that represent their computing capabilities. There are non-functional resources $r_{n,x}$ that offer an amount of computing units to be consumed, where $x$ corresponds to the type of available resource --- CPU ($r_{n,CPU}$), GPU ($r_{n,GPU}$) or memory ($r_{n,MEM}$). Otherwise, there exist functional resources $f_{n,y}$ that consider the capacity to perform concrete functionalities with sensors, actuators or specific hardware that are available, indicated by $y$, such as GPS ($f_{n,GPS})$, video recording ($f_{n,VIDEO}$), etc.

Whenever a direct connection between two nodes exists, a  link $l \in L$ is defined by several features that correspond to the usage and quality of the link. First, the bandwidth capacity $b_l$ that measures the number of available units to allocate data messages. Then, the quality of the link is defined by the one-way propagation delay $d_l$ and the reliability $\rho_l$, which represents the packet delivery ratio, in order to establish a cost to use a single link $\delta_l = d_l /\rho_l$ for unicast routing.

In multi-hop mesh networks, two nodes $n$ and $n'$ can communicate through a subset of links that constitute a specific path $p(n,n') \in P(n,n')$, where $P(n,n')$ is the set of all the paths that provide connectivity between $n$ and $n'$.
The specific path is determined by the routing schema used in the network, and a metric cost $\varepsilon_{p(n,n')}$ is associated to the entire path since it can follow different subpaths.

Table~\ref{tab:substrate_network_framework} summarizes the elements of the substrate network and their representation. 

\begin{table}[!htbp]
\small
\centering
\caption{Elements of the substrate framework}
\label{tab:substrate_network_framework}
\renewcommand{\arraystretch}{1.2}
\begin{tabular}{lp{6cm}}
\toprule
${N}$ & Set of nodes ($n$) in the network topology of the physical infrastructure.\\
${L}$ & Set of links ($l$) in the network topology of the physical infrastructure.\\
${r_{n,x}}$ & Non-functional resources of type $x$ (e.g. CPU) in node $n$.\\
${f_{n,y}}$ & Functional resources of type $y$ (e.g. particular sensor) in node $n$.\\
${b_{l}}$ & Available bandwidth on link $l$.\\
${d_{l}}$ & One way propagation delay on link $l$.\\
${\rho_{l}}$ & Packet delivery ratio on link $l$.\\
${\delta_{l}}$ & Cost to use link $l$.\\
${P(n,n')}$ & Set of paths ($p(n,n')$) that provide connectivity between nodes $n$ and $n'$.\\
${\varepsilon_{p(n,n')}}$ & Cost of the path $p(n,n')$.\\
\bottomrule
\end{tabular}
\end{table}

\subsection{Virtual network}

Similarly, each application is a virtual request represented by a directed graph $\vec{G^v}=(S,C)$, where $S$ is the set of nano-services that need to be executed and $C$ are the communication channels that transport the messages.

Each nano-service $s \in S$ is characterized by a set of requirements that must be fulfilled to operate successfully. On the one hand, non-functional requirements $r^v_{s,x}$ which correspond to the demand of resource units of type $x$ --- CPU ($r^v_{s,CPU}$), GPU ($r^v_{s,GPU}$) or memory ($r^v_{s,MEM}$). On the other hand, the functional ones $f^v_{s,y}$ which are particular service functions associated to sensors, actuators or specific hardware, indicated by $y$, that need to be executed to proceed with the task, e.g. an alarm loudspeaker ($f^v_{s, ALARM}$).

Regarding the communication path, each channel $c \in C$ has a set of attributes that define the resources and quality needed. To deliver a message flow, an amount of bandwidth resource units $b^v_c$ must be reserved and it must guarantee that the maximum delay $d^v_c$ is not exceeded and a minimal reliability $\rho^v_c$ for the channel in order to ensure that the communication cost is not higher than $\delta^v_c = d^v_c / \rho^v_c$.
%
Moreover, channels are intended to connect two single services $c=(c_{src}, c_{dst})$, that is, the communication flow starts from a source service $src(c) = c_{src} \in S$ and it ends in a destination service $dst(c) = c_{dst} \in S$.

Table~\ref{tab:virtual_network_framework} summarizes the elements of the application request that has to be distributed and deployed on the physical infrastructure network. 

\begin{table}[!htbp]
\small
\centering
\caption{Elements of the virtual request (application)}
\label{tab:virtual_network_framework}
\renewcommand{\arraystretch}{1.2}
\begin{tabular}{lp{6cm}}
\toprule
${S}$ & Set of nano-services ($s$) that are requested to be deployed in the network.\\
${C}$ & Set of communication channels ($c$) connecting the nano-services ($c_{src}, c_{dst}$).\\
$r^v_{s,x}$  & Non-functional resources of type $x$ (e.g. CPU) needed by service $s$.\\
$f^v_{s,y}$ & Functional resources of type $y$ (e.g. particular sensor) required by service $s$.\\
${b^v_{c}}$ & Bandwidth required by channel $c$.\\
${d^v_{c}}$ & Maximum delay that channel $c$ supports.\\
${\rho^v_{c}}$ & Minimal reliability required by channel $c$.\\
${\delta^v_{c}}$ & Maximum cost tolerated by channel $c$.\\
\bottomrule
\end{tabular}
\end{table}

It must be noted that dataflow applications are deployed for continuous operation  until a termination instruction is launched. This means that resources are consumed indefinitely. Moreover, services are invoked as message flows are delivered, that is, they depend on previous data received.

\subsection{Problem description} 

The process of embedding virtual resources into a network infrastructure consists of two steps. In the first step, application services must be matched to network nodes considering the computing constraints. In the second step, the most suitable communication path between the chosen nodes must be found, taking advantage of the broadcast features of wireless networks and using an anypath routing mechanism that allows multiple paths to be used simultaneously in order to improve the overall reliability of the wireless channels.

\subsubsection{Service-to-Node Mapping}

each nano-service $s \in S$ from the virtual request $\vec{G^v}$ is assigned to a node $n \in N$ in the substrate network $G^s$ as defined by the mapping function

\begin{equation}
\begin{split}
\mathcal{M}^N: S & \rightarrow N \\
s & \rightarrow \mathcal{M}^N(s)
\end{split}
\end{equation}

\noindent which assigns physical resources to a nano-service satisfying the both following conditions

\begin{equation}
r^v_{s,x} \leq r_{\mathcal{M}^N(s),x}, \forall x \in \{CPU, GPU, MEM\}
\end{equation}

\begin{equation}
\{f^v_{s,y}\} \subseteq \{f_{\mathcal{M}^N(s),y}\}, \forall y
\end{equation}

\noindent Thus, all non-functional resources in the selected physical node $r_{\mathcal{M}^N(s),x}$ must be enough to satisfy the demand $r^v_{s,x}$, and the node is able to perform the functional operations required by the nano-service $f^v_{s,y}$.

\subsubsection{Channel-to-Link Mapping}

for each channel $c \in C$ connecting two nano-services $c_{src}, c_{dst} \in S$, it is necessary to assign a path that connects the nodes that are assigned to the nano-services. We define

\begin{equation}
\begin{split}
    \mathcal{M}^L: C & \rightarrow P(\mathcal{M}^N(c_{src}), \mathcal{M}^N(c_{dst})) \\
    c & \rightarrow \mathcal{M}^L(c) = p(\mathcal{M}^N(c_{src}), \mathcal{M}^N(c_{dst}))
\end{split}
\end{equation}

\noindent which establishes the path satisfying that the bandwidth $b_{l_p}$ of each link $l_p \in L$ constituting the path is equal or higher than the bandwidth required by the channel

\begin{equation}
    \label{eq:bandwidthmin}
    b_{l_p} \geq b^v_c, \forall l_p \in p(\mathcal{M}^N(c_{src}), \mathcal{M}^N(c_{dst})) \\
\end{equation}

\noindent and meeting also the restriction in terms of the maximum acceptable cost for the quality of the communication as follows

\begin{equation}
    \label{eq:costeruta}
    \varepsilon_{p(\mathcal{M}^N(c_{src}), \mathcal{M}^N(c_{dst}))} \leq \delta^v_c
\end{equation}

\begin{figure}[!ht]
    \centering
    \includegraphics[width=0.6\linewidth]{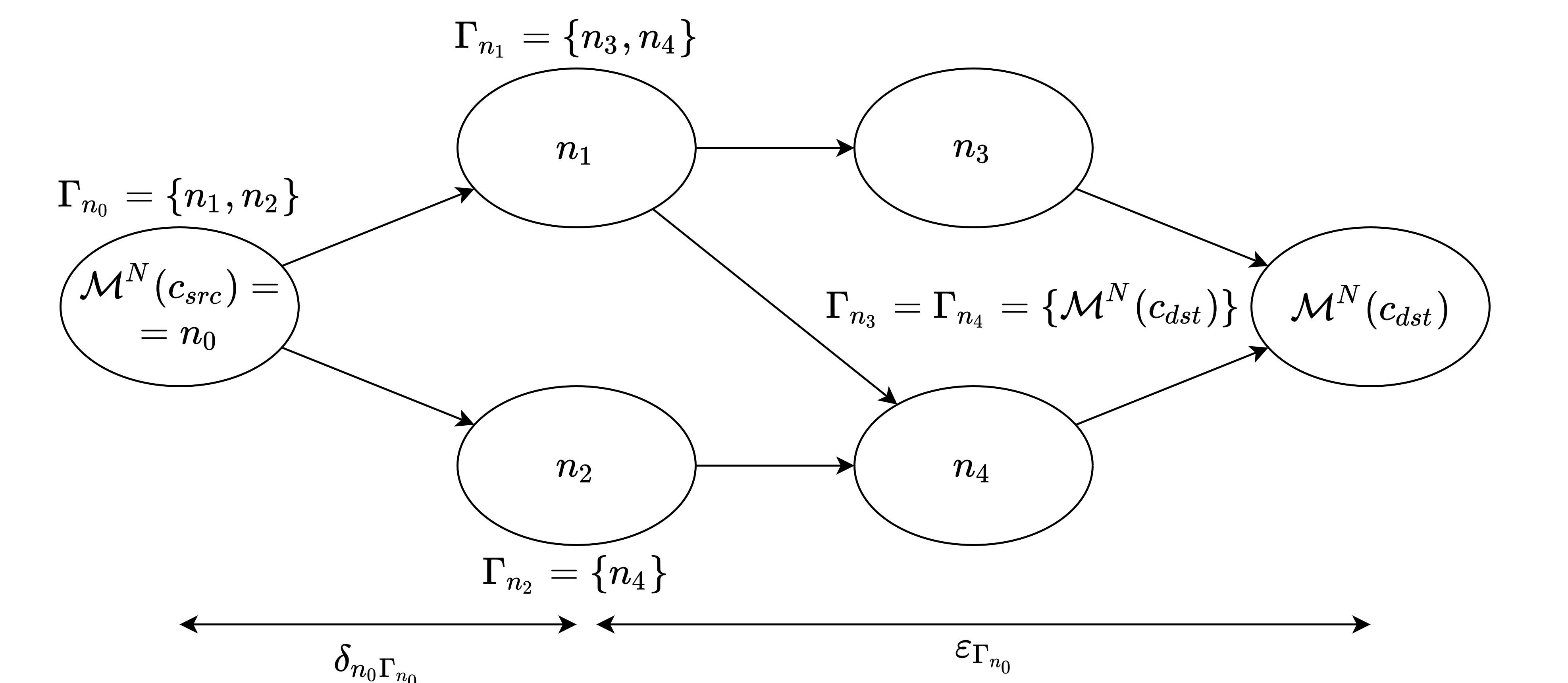}
    \caption{Sequence of hyperlinks in anypath routing.}
    \label{fig:anypath_hyperlink}
\end{figure}

It is important to note that $\mathcal{M}^L$ establishes the path $p(\mathcal{M}^N(c_{src}), \mathcal{M}^N(c_{dst}))$ according to the anypath routing technique \cite{multirate_anypath}. Data packets are forwarded from the source $\mathcal{M}^N(c_{src})$ to the destination $\mathcal{M}^N(c_{dst})$ by relaying the message from a transmitter $n_i \in N$, starting from $i=0$, to any of the corresponding receivers $n_i \in \Gamma_{n_i} \subset N$ as shown in Figure \ref{fig:anypath_hyperlink} recursively to reach the complete anypath route. It is important to note that $\mathcal{M}^L$ establishes the path according to the anypath routing technique. This approach selects the path $p(\mathcal{M}^N(c_{src}), \mathcal{M}^N(c_{dst}))$ that uses all the links that connect the source $\mathcal{M}^N(c_{src})$ and the destination  $\mathcal{M}^N(c_{dst})$ and satisfy the restrictions in expressions\ref{eq:bandwidthmin} and \ref{eq:costeruta}. Then, data packets are forwarded from each node $n_i \in N$ to any of the next set of nodes that are part of the path $p$. We denominate those receivers as $\Gamma_{n_{i}}$ (for node $n_i$), which is a subset of $N$.

Each step from a node in the anypath route can therefore be viewed as a hyperlink $(n_{i},\Gamma_{n_{i}})$ ordered by the chosen routing priority mechanism. The hyperlink is characterized by new quality parameters that represent the simultaneous usage of each sublink, thus it offers new delivery probability $\rho_{n_{i}\Gamma_{n_i}}$, proportional to delivering correctly the message to any of the hyperlink nodes, and communication delay $d_{n_{i}\Gamma_{n_i}}$, which is the maximum delay of all the links in the hyperlink  because the message may be potentially delivered to any of the nodes in the hyperlink.

\begin{equation}
    \rho_{n_{i}\Gamma_{n_{i}}} = 1 - \prod_{m \in \Gamma_{n_{i}}} (1 - \rho_{{n_i}m})
\end{equation}

\begin{equation}
    d_{n_{i}\Gamma_{n_i}} = \max_{m \in \Gamma_{n_i}}(d_{n_{i}m})
\end{equation}

Accordingly, the cost $\delta_{n_{i}\Gamma_{n_i}}$ of using a single hyperlink is proportional to its latency $d_{n_{i}\Gamma_{n_i}}$, and the \textit{expected transmissions} (ETX) \cite{multi-hop_wireless_metric} $1/r_{n_{i}\Gamma_{n_i}}$ to send correctly a message in any of its sublinks in average.

\begin{equation}
    \delta_{n_{i}\Gamma_{n_i}} = \frac{d_{n_{i}\Gamma_{n_i}}}{\rho_{n_{i}\Gamma_{n_i}}}
\end{equation}

Furthermore, the entire virtual path from a node $n \in N$ to a given destination is characterized by the \textit{expected anypath transmission time} (EATT) metric, which is derived from the combination of the \textit{expected transmission time} (ETT) \cite{routing_multi-radio_multi-hop_mesh} and the \textit{expected anypath transmissions} (EAT) \cite{opportunistic_anypath}. It can be calculated recursively for each forwarding set as follows:

\begin{equation}
    \varepsilon_{n_i}=\delta_{n_{i}\Gamma_{n_i}} + \varepsilon_{\Gamma_{n_i}}
\end{equation}

\noindent which takes, on the one side, the cost to use the hyperlink $\delta_{n_{i}\Gamma_{n_i}}$ and, on the other side, the cost to send a message from the hyperlink to the final destination as indicated hereunder:

\begin{equation}
    \varepsilon_{\Gamma_{n_i}} = \sum_{m \in \Gamma_{n_i}} w_{m} \varepsilon_{m}
\end{equation}

\noindent where $w_{m}$ weights the remaining cost of the following sub-hyperlinks $\varepsilon_{m}$ according to the probability of each node in the forwarding set to be the actual anypath forwarder according to their priority position in the hyperlink and defined as follows:

\begin{equation}
    w_{m} = \frac{\rho_{n_{i}m}\prod_{k=1}^{m-1}(1 - \rho_{n_{i}k})}{\rho_{n_{i}\Gamma_{n_i}}}
\end{equation}

In an opportunistic routing scheme, as in the case of anypath routing, a packet may be correctly received by more than one node in the forwarding set due to broadcast mode, but only one can relay it to follow the path to the final destination. Thus, it is necessary to set up a priority mechanism at the link layer that enforces a single receiver with the highest priority to forward the packet, and so on.

Table~\ref{tab:algorithm_framework} summarizes the elements used by $\mathcal{M}^L$ for the creation of the anypath route.

\begin{table}[!htbp]
\small
\centering
\caption{Elements used for the establishment of a path}
\label{tab:algorithm_framework}
\renewcommand{\arraystretch}{1.2}
\begin{tabular}{lp{6cm}}
\toprule
$\mathcal{M}^N$ & Function assigning services to nodes. \\
$\mathcal{M}^L$ & Function assigning channels to paths. \\
$\rho_{n_{i}\Gamma_{n_i}}$ &  Delivery probability through the anypath receivers ($\Gamma_{n_i}$) for node $i$. \\
$d_{n_i\Gamma_{n_i}}$ & Maximum delay of all the links of the anypath route from node $i$. \\
$ \delta_{n_i\Gamma_{n_i}}$ & Cost of using the anypath route from node $i$. \\ 
$\varepsilon_{n_i}$ &  Expected anypath transmission time from node $i$. \\ 
$\varepsilon_{\Gamma_{n_i}}$ &  Cost to send a message from the potential receivers of the node $i$ to the final destination. \\
$w_{m}$ & Weight of the remaining cost of the following sup-hyperlinks. \\ 

\bottomrule
\end{tabular}
\end{table}

\subsection{Objective}

The main idea of this work is to design a resource allocation mapping algorithm that maximizes the usage of the available physical infrastructure through the embedding of as many VN requests as possible to execute them simultaneously. Drawing on the nature of wireless communications, anypath routing is exploited to obtain virtual communication channels that could provide the needed communication quality requirements.

In order to rigorously evaluate the performance of the proposed QRPAD-VNE algorithm, as well as making design decisions about it, we have formulated standard metrics of virtual resource allocation. It is based on the set of total VN requests that the system has received $\mathit{VNR}_{\mathit{Tot}}$, the amount of them that have been accepted $\mathit{VNR}_{\mathit{Acc}}$ to be allocated, and the blocked ones $\mathit{VNR}_{\mathit{Blo}}$ due to the lack of available resources.

\subsubsection{Acceptance Ratio/Blocking Ratio}

it indicates the fraction of VN requests that have been accepted to be embedded $\mathit{VNR}_{\mathit{Acc}}$ over the total ones $\mathit{VNR}_{\mathit{Tot}}$ received by the system.

\begin{equation}
    \parbox{1.8cm}{\centering $\mathit{Acceptance}$\\$\mathit{Ratio}$} = \frac{|\mathit{VNR}_{\mathit{Acc}}|}{|\mathit{VNR}_{\mathit{Tot}}|}
\end{equation}

Similarly, the ratio of refused ones can be calculated from the number of VN requests that could not be allocated $\mathit{VNR}_{\mathit{Blo}}$ in the substrate network due to lack of available resources.

\begin{equation}
    \parbox{1.5cm}{\centering $\mathit{Blocking}$\\$\mathit{Ratio}$} = \frac{|\mathit{VNR}_{\mathit{Blo}}|}{|\mathit{VNR}_{\mathit{Tot}}|} = 1 - \parbox{1.8cm}{\centering $\mathit{Acceptance}$\\$\mathit{Ratio}$}
\end{equation}

\begin{equation}
\begin{gathered}
    \mathit{VNR}_{\mathit{Tot}} = \mathit{VNR}_{\mathit{Acc}} \cup \mathit{VNR}_{\mathit{Blo}} \\
    \mathit{VNR}_{\mathit{Acc}} \cap \mathit{VNR}_{\mathit{Blo}} = \emptyset
\end{gathered}
\end{equation}

\subsubsection{Revenue}

it measures the total amount of resources that are requested by a single VN request $\vec{G^v}$ in terms of nano-service $s \in S$ and channel $c \in C$ demand. It is used to model the ``theoretical price'' of using all computing $r^v_{n,x}$ and network $b^v_c$ resources. Despite that in many of the scenarios for FANETs we do not expect having operators expecting an economic benefit, we use such ``theoretical price'' as a tool for optimizing the usage of the network.  Note that computing, network resources, and communication requirements have different nature and units, thus requiring  weighing each kind of revenue element in accordance of their importance and usage cost, by using coefficients $\alpha_x$ and $\beta$ in each addend of equation \ref{eq:revenue} for node and link resources, respectively.

\begin{equation}
\label{eq:revenue}
    R(\vec{G^v})=\sum_{x}\alpha_{x}\sum_{s}r^v_{s,x} + \beta\sum_{c}b^v_c
\end{equation}

In view of the complete demand of all embedded VN requests $\vec{G^v} \in \mathit{VNR}_{\mathit{Acc}}$, the potential revenue to obtain in the whole network is defined by expression \ref{eq:embrevenue}.

\begin{equation}
    \label{eq:embrevenue}
    \parbox{1.8cm}{\centering $\mathit{Emb.}$\\$\mathit{Revenue}$} = \sum_{\vec{G^v} \in \mathit{VNR}_{\mathit{Acc}}} R(\vec{G^v}) = \sum R(\mathit{VNR}_{\mathit{Acc}})
\end{equation}

\subsubsection{Cost}

it computes the number of physical resources that are actually used to embed a single VN request $\vec{G^v}$ regarding node $n \in N$ and link $l \in L$ usage. Specifically, the operator measures how much of the available resources has to reserve for the corresponding service $r^v_{s,x}$ and how many bandwidth $b^v_c$ is requested to the physical links $l_{p(c_{src}, c_{dst})}$ of the virtual path $p(c_{src}, c_{dst})$. In addition, coefficients $\alpha_{x}'$ and $\beta'$ must be applied to give each resource element the actual cost.

\begin{equation}
    C(\vec{G^v})=\sum_{x}\alpha_{x}'\sum_{s}r^v_{s,x} + \beta'\sum_{c} b^v_c \cdot \vert l_{p(c_{src}, c_{dst})}\vert
\end{equation}

The embedding of all accepted VN requests $\vec{G^v} \in \mathit{VNR}_{\mathit{Acc}}$ has a total cost for the network operator, which is the sum of all individual costs.

\begin{equation}
    \parbox{1.2cm}{\centering $\mathit{Emb.}$\\$\mathit{Cost}$} = \sum_{\vec{G^v} \in \mathit{VNR}_{\mathit{Acc}}} C(\vec{G^v}) = \sum C(\mathit{VNR}_{\mathit{Acc}})
\end{equation}

\subsubsection{Revenue-Cost Relation}

it indicates the performance of the embedding algorithm for accepted VN requests.

\begin{equation}
    R/C = \frac{\mathit{Embedding\ Revenue}}{\mathit{Embedding\ Cost}}
\end{equation}

In the case of same respective weight coefficients in both Revenue and Cost, this ratio can be regarded as a normalized efficiency metric which varies from 0 to 1, this latter case when virtual links are mapped to single physical links.

Hence, we adopt a conscious validation framework that guarantees comprehensive performance comparison of different VNE patterns in terms of resource exploitation. To sum up, the specific goals are maximizing Revenue-Cost relation through minimizing the Cost of usage of physical infrastructure, while having simultaneously a high Acceptance Ratio that avoids fragmentation to prevent rejecting potential applications. In designing the QRPAD-VNE algorithm, these metrics are borne in mind and some actions are taken, such as prioritizing high-demand requests, choosing short and highly reliable communication paths or shaping network data rates restricted to the required demand.

Table~\ref{tab:metrics} summarizes the elements used for the evaluation of the framework. 

\begin{table}[!htbp]
\small
\centering
\caption{Evaluation framework elements}
\label{tab:metrics}
\renewcommand{\arraystretch}{1.2}
\begin{tabular}{p{1.4cm}p{6.5cm}}
\toprule
$\mathit{VNR}_{\mathit{Tot}}$ & Number of $\mathit{total}$ VN requests in the system. \\ 
$\mathit{VNR}_{\mathit{Acc}}$ & Number of $\mathit{accepted}$ VN requests in the system. \\ 
$\mathit{VNR}_{\mathit{Blo}}$ & Number of $\mathit{bloqued}$ VN requests in the system. \\ 
$\mathit{Acceptance}$ $\mathit{Ratio}$ & Ratio between $\mathit{accepted}$ and $\mathit{total}$ requests. \\ 
$\mathit{Blocking}$ $\mathit{Ratio}$ &  Ratio between $\mathit{bloqued}$ and $\mathit{total}$ requests. \\ 
$R(\vec{G^v})$ & Theoretical revenue obtained from the virtual request $\vec{G^v}$.\\ 
$C(\vec{G^v})$ & Cost of using the resources required to implement the virtual request $\vec{G^v}$. \\ 
$R/C$ & Performance of the embedding algorithm for accepted VN requests. If weight and cost use the same coefficients, represents the $\mathit{efficiency}$. \\                 
\bottomrule
\end{tabular}
\end{table}

\section{QRPAD-VNE Algorithm}\label{algorithm}

This section is intended to present a heuristic Quality-Revenue Paired Anypath Dataflow VNE algorithm that solves the abovementioned VNE problem, which is widely known to be NP-hard \cite{vne_complexity}. According to the depicted network and application framework, it proceeds in three embedded stages. On the top level, (A) the management of all the available VN requests in a period of time is performed, then (B) the complete embedding of each single VN request choosing the deployment location of services, and lastly (C) the actual embedding of the communication channel from the specified source and destination services using the anypath routing. In addition, a study about the complexity of the heuristic proposal and an illustrative example are outlined.

\subsection{Window Assignment}

The VNE technology is intended to handle multiple applications concurrently in the best possible manner to successfully allocate them and minimize the number of requests that are rejected due to combinatorial lack of resources. Thereby, we must implement a top level management service, described in Algorithm \ref{alg:window}, that receives VN requests $VNR$ for a time window $\mathcal{W}$ (\textsc{Collect}) in order to find the correct embedding (\textsc{Embed}) and deploy them, but first they are ordered (\textsc{Sort}) according to a variation of the Revenue metric, defined as Quality-Revenue (\textsc{Quality-Revenue}) in Equation \ref{eq:quaity_revenue_metric} for each request $\vec{G^v}$, giving priority to the bigger ones.

\begin{equation} \label{eq:quaity_revenue_metric}
    R^q(\vec{G^v})=R(\vec{G^v}) + \gamma\sum_{c}\frac{\rho^v_c}{d^v_c}
\end{equation}

\begin{algorithm}
\caption{Window}\label{alg:window}
\begin{algorithmic}[1]
\Require $G^s, \mathcal{W}$

\State $\mathit{VNR} \gets$ \Call{Collect}{$\mathcal{W}$}
\For{\textbf{each} $\vec{G^v} \in \mathit{VNR}$}
\State $R_{\mathit{VNR}} =$ \Call{Quality-Revenue}{$\vec{G^v}$}
\EndFor
\State $\mathit{VNR}\prime \gets$ \Call{Sort}{$\mathit{VNR}$, $R_{\mathit{VNR}}$, $``descend"$}

\For{\textbf{each} $\vec{G^v} \in \mathit{VNR}\prime$}
\State $G^s , \mathcal{M}^N(S), \mathcal{M}^L(C) \gets$ \Call{Embed}{$\vec{G^v}$, $G^s$}
\EndFor

\end{algorithmic}
\end{algorithm}

In this way, we try to allocate the greater number of applications at the same time so that we exploit resources optimally and obtain the best profit of them.

\subsection{Communication Quality Revenue-Pair VNE Algorithm}

The major contribution of this work is to allocate a specific dataflow application $\vec{G^v}$ into the given wireless substrate network $G^s$, so we have developed the Quality-Revenue Paired Anypath Dataflow Virtual Network Embedding in Algorithm \ref{alg:vne}, that process each application, with its pair of services and the communication channel $(c, c_{src}, c_{dst})$, regarding the requested resources.

In order to do so, each pair of services processing information transmitted through the channel (flow pairs) are considered as virtual sub-graphs $\vec{G^v_c} = (\{c_{src}, c_{dst}\}, {c})$ and sorted according to the Quality-Revenue metric (\textsc{Quality-Revenue}) as follows:

\begin{equation}
    R^q(\vec{G^v_c})=R(\vec{G^v_c}) + \gamma\frac{\rho^v_c}{d^v_c}
\end{equation}

\noindent where the actual revenue $R(\vec{G^v_c})$, which in this specific case corresponds to two nano-services $c_{src}, c_{dst}$ and a single channel $c$, adds the communication quality factor defined by the minimum required reliability $\rho^v_c$ and the maximum allowed delay $d^v_c$ of the communication channel and a coefficient $\gamma$ that weights it to emphasize the communication quality requirements regarding the revenue value.

Then, the list of sub graphs sorted in descending order (\textsc{Sort}) is used to allocate them independently to find a mapping association that satisfies the service and channel requirements using a virtual anypath route (see section \ref{sec:anypath}). For every channel $c$, the procedure is performed in three stages, firstly the destination node $n_{dst}$ from which the anypath routing execution is chosen, secondly the anypath costs $\varepsilon$ and routes $F$ from the rest of nodes are calculated (\textsc{Anypath}), and thirdly the source node $n_{src}$ is guessed from the potential nodes $N_{src}$ that comply with the functional and non-functional requirements and establishes the virtual path with the lowest number of necessary links (\textsc{Min-Number-Links}), in order to finally allocate properly the source $c_{src}$ and destination $c_{dst}$ services linked by the corresponding virtual path $\mathcal{M}^L(c)$ while ensuring that the cost of embedding is minimized.

The algorithm must consider four alternatives according to the allocation state of source $c_{src}$ and destination $c_{dst}$ services of the current channel $c$: 

\begin{itemize}
    \item \ul{Neither source service $c_{src}$ nor destination service $c_{dst}$ are yet allocated:} since none of them has collided in a previous iteration, it is first necessary to select the destination node $n_{dst}$ to run the anypath routing, which will be from the suitable nodes the one with highest local PDR (\textsc{Max-PDR}), because it is preferable to have low loss ratio in the final hop to avoid complete and costly dataflow retransmission, and the node resources are modified to ensure consistency (\textsc{Update-Node}). In addition, the potential source nodes $N_{src}$ that are suitable for the source service $c_{src}$ must be gathered (\textsc{Suitable}) to filter which virtual paths must be considered. Once the virtual paths are known, the one with the lowest number of physical links used and that meets the channel quality requirements is chosen, the mapping update is proceed, otherwise VNE is aborted. At last, the graph is updated according to the embedding done and information about it is reported to the global environment to deploy the virtual request.
    \item \ul{Destination service $c_{dst}$ already allocated, but source service $c_{src}$ not:} the same operation as in the previous point, but taking into account that the destination node $n_{dst}$ is already known because the destination service $c_{dst}$ was assigned previously.
    \item \ul{Source service $c_{src}$ already allocated, but destination service $c_{dst}$ not:} in this case the source service $c_{src}$ is assigned to a node, so the underlying logic is similar to the previous one, but reversely since communication links are considered symmetrical. The destination node $n_{dst}$ will be the one already assigned to the source service $c_{src}$ and the potential source nodes $N_{src}$ will take into account the destination service $c_{dst}$ requirements. This way, it is only necessary to invert the virtual path direction and to apply the same strategy.
    \item \ul{Both source $c_{src}$ and destination $c_{dst}$ services already allocated:} only the communication channel must be embedded to the virtual path. Then, taking the destination node $n_{dst}$ of the respective destination service $c_{dst}$ and considering that the only potential source node $N_{src}$ is the one already assigned to the source service $c_{src}$, we can validate if the resulting anypath route meets the conditions. If this is the case, the channel embedding is achieved.
\end{itemize}

Furthermore, it is important to clarify a relevant issue of the algorithm: the way anypath routes are discovered. The execution is actually restricted to a undirected subgraph $G^s\prime$ that only contains the links with sufficient bandwidth resources for the corresponding channel bandwidth $b^v_c$ (\textsc{Subgraph}). Through this, bottlenecks are avoided and the resulting subgraph $G^s\prime$ is then pruned to the chosen destination node $c_{dst}$ to obtain the potential anypath routes $F$ as explained in section \ref{sec:anypath}.

Apart from choosing a route that exceeds the channel cost, the embedding execution may be faulty terminated because of the lack of resources about suitable nodes for the services or isolated nodes by inactive low-bandwidth links.

\begin{algorithm}
\caption{QRPAD-VNE (\textsc{Embed})}\label{alg:vne}
\begin{algorithmic}[1]

\Require $G^s = (N, L), \vec{G^v} = (S, C)$


\For{\textbf{each} $c \in C$}
\State $\vec{G^v_c} \gets (\{c_{src}, c_{dst}\}, \{c\})$
\State $R^q_c \gets$ \Call{Quality-Revenue}{$\vec{G^v_c}$}
\EndFor
\State $C\prime \gets$ \Call{Sort}{$C$, $R^q$, ``descend"}

\For{\textbf{each} $c \in C\prime$}

\If{$c_{src}$ not allocated AND $c_{dst}$ not allocated}

\State $n_{dst} \gets$ \Call{Max-PDR}{$c_{dst}$, $N$}
\State $N \gets$ \Call{Update-Node}{$N$, $n_{dst}$, $c_{dst}$}
\State $N_{src} \gets$ \Call{Suitable}{$c_{src}$, $N$}

\ElsIf{$c_{src}$ not allocated AND $c_{dst}$ allocated}

\State $n_{dst} \gets \mathcal{M}^N(c_{dst})$
\State $N_{src} \gets$ \Call{Suitable}{$c_{src}$, $N$}

\ElsIf{$c_{src}$ allocated AND $c_{dst}$ not allocated}

\State $n_{dst} \gets \mathcal{M}^N(c_{src})$
\State $N_{src} \gets$ \Call{Suitable}{$c_{dst}$, $N$}

\ElsIf{$c_{src}$ allocated AND $c_{dst}$ allocated}

\State $n_{dst} \gets \mathcal{M}^N(c_{dst})$
\State $N_{src} \gets \{\mathcal{M}^N(c_{src})\}$

\EndIf

\State $G^s\prime \gets$ \Call{Subgraph}{$G^s$, $b^v_c$}

\State $\vec{G^s\prime} \gets$ \Call{Prune}{$G^s\prime$, $n_{dst}$}
\State $F, \varepsilon \gets$ \Call{Anypath}{$\vec{G^s\prime}$, $n_{dst}$}

\State $N_{src}\prime \gets n \in N_{src}\ /\ \varepsilon_n \leq d(c)/r(c)$
\If{$N_{src}\prime$ is empty}
\State report ERROR
\EndIf
\State $n_{src} \gets$ \Call{Min-Number-Links}{$F$, $N_{src}\prime$}

\If{$c_{src}$ not allocated AND $c_{dst}$ not allocated}

\State $N \gets$ \Call{Update-Node}{$N$, $n_{src}$, $c_{src}$}
\State $\mathcal{M}^N(c_{src}) \gets n_{src}$
\State $\mathcal{M}^N(c_{dst}) \gets n_{dst}$

\ElsIf{$c_{src}$ not allocated AND $c_{dst}$ allocated}

\State $N \gets$ \Call{Update-Node}{$N$, $n_{src}$, $c_{src}$}
\State $\mathcal{M}^N(c_{src}) \gets n_{src}$

\ElsIf{$c_{src}$ allocated AND $c_{dst}$ not allocated}

\State $N \gets$ \Call{Update-Node}{$N$, $n_{src}$, $c_{dst}$}
\State $\mathcal{M}^N(c_{dst}) \gets n_{src}$
\State $F \gets F^T$

\EndIf

\State $\mathcal{M}^L(c) \gets F_{\mathcal{M}^N(c_{src})}$
\State $L \gets$ \Call{Update-Links}{$L$, $\mathcal{M}^N(c_{src})$, $\mathcal{M}^N(c_{dst})$, $\mathcal{M}^L(c)$, $b^v_c$}

\EndFor

\State \Return $G^s, \mathcal{M}^N(S), \mathcal{M}^L(C)$

\end{algorithmic}
\end{algorithm}

\subsection{Anypath Routing}
\label{sec:anypath} 

The foundation of this routing strategy lies in the capability to define virtual paths to a destination node $n_{dst}$ in a multi-hop wireless network taking advantage of its broadcast feature. It considers a directed tree, whose root is the destination $dst \in N$, but in order to avoid redundant forwarders that may increment the cost of virtual paths, and reduce the number of potential link bandwidth resources to be consumed, it is necessary to prune the substrate network $G^s$ to a directed substrate network $\vec{G^s}$ according to the distance/cost of each node to the chosen destination $dst$ (\textsc{Dist}). In Algorithm \ref{alg:prune}, we divide the undirected links $\vec{L}$ in two directed links (\textsc{Undirected-2-Directed}) and eliminate those which depart from the nearest node $(n, n')$.

\begin{algorithm}
\caption{Prune Graph (\textsc{Prune})}\label{alg:prune}
\begin{algorithmic}[1]
\Require $G^s, dst$

\State $\vec{L} \gets$ \Call{Undirected-2-Directed}{$L$}
\For{\textbf{each} $l=(n,n') \in L$}
\If{\Call{Dist}{$n$, $dst$} $\leq$ \Call{Dist}{$n'$, $dst$}}
\State $\vec{L} \gets \vec{L} - \{(n,n')\}$
\EndIf
\EndFor

\State $\vec{G^s} \gets \{N,\vec{L}\}$
\State \Return $\vec{G^s}$

\end{algorithmic}
\end{algorithm}

Then, we compute the anypath routes of all nodes of $\vec{G^s}$ (the resulting network after applying the pruning algorithm) backward from the specified node for the destination service ($dst$) following the Algorithm \ref{alg:anypath} that calculates the metric cost according to the problem statement.

It iterates all the nodes $n' \in N$ of the graph, starting from the destination $dst \in N$ and ending at the farthest one (\textsc{Extract-Min}), taking all incoming edges $\vec{l} \in I$ from the neighbors $n \in N$ (\textsc{Incoming-Links}) to ascertain whether the running node $n'$ must be considered as potential next-hop in the virtual path, if so $n'$ is included in the corresponding hyperlink set of the neighbor $\Gamma_n$ and its cost $\varepsilon_n$ is calculated using the current distance to the destination in that iteration, to order the priority at the hyperlinks. 

The algorithm returns the anypath routes in $F$, that contains the specific hyperlinks $F_n$ for each single node $n$ for implementing the communication network of the virtual path to the destination $dst$, together with their corresponding costs $\varepsilon_n$ for each node $n$ in the set $\varepsilon$.

\begin{algorithm}
\caption{Anypath Routing (\textsc{Anypath})}\label{alg:anypath}
\begin{algorithmic}[1]
\Require $\vec{G^s}, dst$

\For{each $n \in N$}
\State $\varepsilon_n \gets \infty$
\State $F_n \gets \emptyset$
\EndFor

\State $\varepsilon_{dst} \gets 0$
\State $S \gets \emptyset$
\State $Q \gets N$

\While{$Q \neq \emptyset$}

\State $n' \gets$ \Call{Extract-Min}{$Q$}
\State $S \gets S \cup \{n'\}$

\State $I \gets$ \Call{Incoming-Links}{$n'$}
\If{$I = \emptyset$}
\State break
\EndIf

\For{each $\vec{l} = (n, n') \in I$}

\State $\Gamma_n \gets F_{n} \cup {n'}$
\If{$\varepsilon_{n} > \varepsilon_{n'}$}
\State $\varepsilon_{n} \gets \delta_{n\Gamma_{n}} + \varepsilon_{\Gamma_n}$
\State $F_{n} \gets \Gamma_{n}$
\EndIf

\EndFor

\EndWhile

\State \Return $F, \varepsilon$

\end{algorithmic}
\end{algorithm}

\subsection{Complexity}

The proposed algorithmic mapping strategy, although not being optimal, poses an effective alternative to achieve a solution to the extremely challenging NP-hardness of the addressed VNE problem. Hereafter a complexity study of the cost to handle a single VN request allocation run is detailed.

As exposed in Algorithm \ref{alg:vne}, a VN request is processed per communication channel, which depend on several processes \textsc{Quality-Revenue}, \textsc{Sort}, \textsc{Suitable}, \textsc{Max-PDR}, \textsc{Subgraph}, \textsc{Prune}, \textsc{Anypath}, \textsc{Min-Number-Links} and \textsc{Update-Links} that operate on non-constant data structures.

On the one side, \textsc{Quality-Revenue} and \textsc{Sort} functions operate over $|C|$ channels then a complexity of $\mathcal{O}(|C|)$ is associated, \textsc{Suitable}  and \textsc{Max-PDR} functions have a complexity of $\mathcal{O}(|N|)$ since they target the $|N|$ nodes of the substrate graph, and \textsc{Subgraph}, \textsc{Update-Links} and \textsc{Min-Number-Links} functions have a complexity of $\mathcal{O}(|L|)$ since they do the same but on the $L$ substrate links. On the other side, \textsc{Prune} and \textsc{Anypath} functions are the essence of the single-rate anypath routing explained in \cite{multirate_anypath}, which demonstrates that the complexities are both $\mathcal{O}(|N| \cdot log(|N|))$.

Therefore, the overall complexity of the QRPAD-VNE algorithm is $\mathcal{O}(|C| \cdot (|N| \cdot log(|N|) + |L|))$, because it iterates through $|C|$ virtual communication channels and $\mathcal{O}(|N| \cdot log(|N|)) > \mathcal{O}(|N|)$.

\subsection{Illustrative Example}

\begin{figure*}[!htb]
    \centering
    \includegraphics[width=\linewidth]{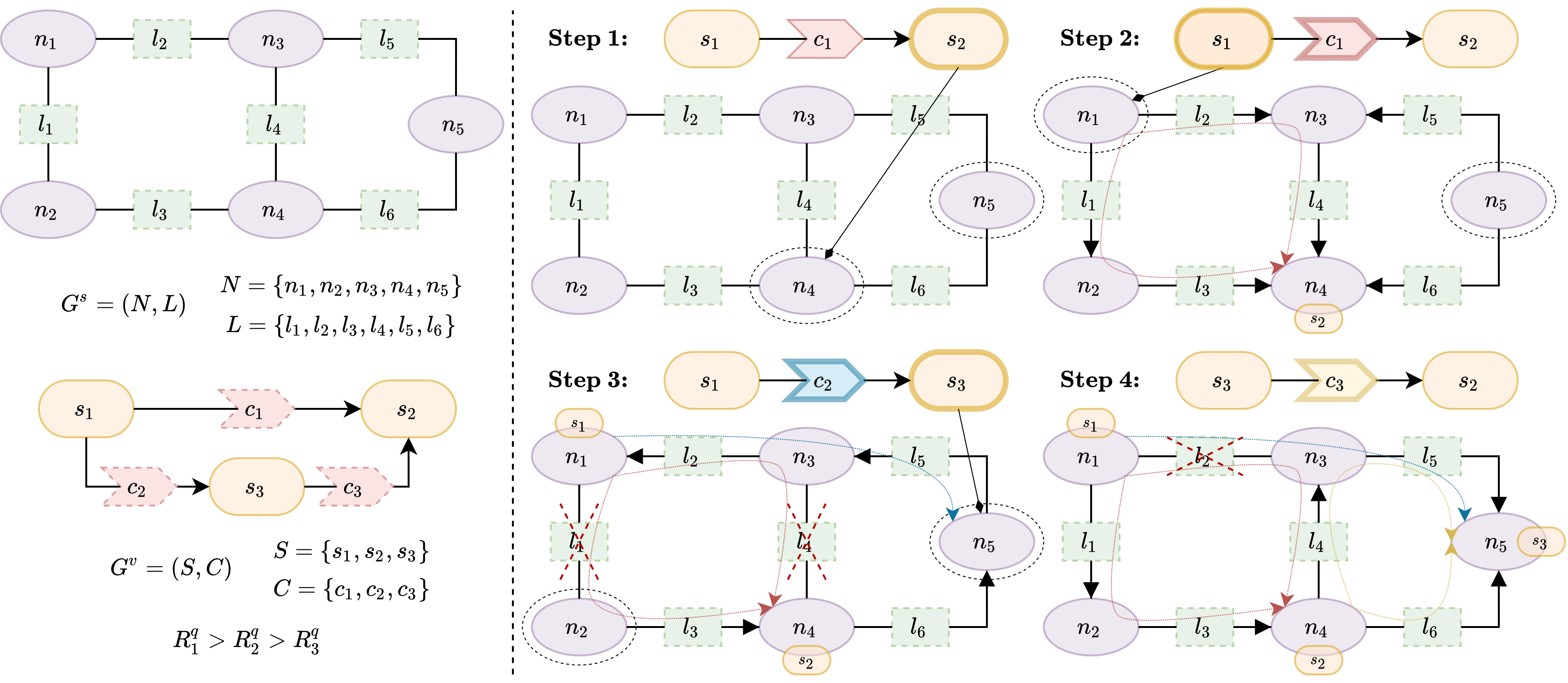}
    \caption{QRPAD-VNE algorithm example}
    \label{fig:vne_example}
\end{figure*}

In order to have a thorough understanding of the QRPAD algorithm, we provide the example in Figure \ref{fig:vne_example}. We proceed with the substrate network $G^s$,  whose node $N$ and link $L$ properties are  specified in Tables \ref{tab:nodes} and \ref{tab:links}, respectively, to allocate the VN request $\vec{G^v}$ according to the requirements in Tables \ref{tab:services} and \ref{tab:channels} for the services $S$ and channels $C$, respectively. Besides, it has to be noted that only non-functional requirements have been considered for the sake of simplicity.

\begin{table}[!htb]
\caption{Node attributes of the VNE example}
\scalebox{0.85}{
\begin{tabular}{|c|c|c|c|cccc}
\cline{1-4} \cline{6-8}
\textbf{Node} & \textbf{CPU} & \textbf{GPU} & \textbf{MEM} & \multicolumn{1}{c|}{} & \multicolumn{1}{c|}{\textbf{CPU}} & \multicolumn{1}{c|}{\textbf{GPU}} & \multicolumn{1}{c|}{\textbf{MEM}} \\ \cline{1-4} \cline{6-8} 
$n_1$ & 50 & 20 & 30 & \multicolumn{1}{c|}{- Step 2 $\rightarrow$} & \multicolumn{1}{c|}{0} & \multicolumn{1}{c|}{0} & \multicolumn{1}{c|}{0} \\ \cline{1-4} \cline{6-8} 
$n_2$ & 20 & 20 & 50 &  &  &  &  \\ \cline{1-4}
$n_3$ & 10 & 10 & 10 &  &  &  &  \\ \cline{1-4} \cline{6-8} 
$n_4$ & 10 & 30 & 30 & \multicolumn{1}{c|}{- Step 1 $\rightarrow$} & \multicolumn{1}{c|}{0} & \multicolumn{1}{c|}{0} & \multicolumn{1}{c|}{10} \\ \cline{1-4} \cline{6-8} 
$n_5$ & 20 & 10 & 50 & \multicolumn{1}{c|}{- Step 3 $\rightarrow$} & \multicolumn{1}{c|}{10} & \multicolumn{1}{c|}{10} & \multicolumn{1}{c|}{0} \\ \cline{1-4} \cline{6-8}
\end{tabular}
}
\centering
\label{tab:nodes}
\end{table}

\begin{table}[!htb]
\caption{Link attributes of the VNE example}
\scalebox{0.85}{
\begin{tabular}{|c|c|c|c|c|c|cc}
\cline{1-4} \cline{6-6} \cline{8-8}
\textbf{Link} & $\mathbf{d_l}$ & $\mathbf{\rho_l}$ & $\mathbf{b_l}$ & \textbf{} & $\mathbf{b_l}$ & \multicolumn{1}{c|}{\textbf{}} & \multicolumn{1}{c|}{$\mathbf{b_l}$} \\ \cline{1-4} \cline{6-6} \cline{8-8} 
$l_1$ & 10 & 0.9 & 70 & $-$ Step 2 $\rightarrow$ & 20 &  &  \\ \cline{1-4} \cline{6-6} \cline{8-8} 
$l_2$ & 10 & 0.9 & 80 & $-$ Step 2 $\rightarrow$ & 30 & \multicolumn{1}{c|}{$-$ Step 3 $\rightarrow$} & \multicolumn{1}{c|}{0} \\ \cline{1-4} \cline{6-6} \cline{8-8} 
$l_3$ & 10 & 0.9 & 100 & $-$ Step 2 $\rightarrow$ & 50 &  &  \\ \cline{1-4} \cline{6-6} \cline{8-8} 
$l_4$ & 10 & 0.9 & 70 & $-$ Step 2 $\rightarrow$ & 20 & \multicolumn{1}{c|}{$-$ Step 4 $\rightarrow$} & \multicolumn{1}{c|}{10} \\ \cline{1-4} \cline{6-6} \cline{8-8} 
$l_5$ & 20 & 0.75 & 100 & $-$ Step 3 $\rightarrow$ & 70 & \multicolumn{1}{c|}{$-$ Step 4 $\rightarrow$} & \multicolumn{1}{c|}{60} \\ \cline{1-4} \cline{6-6} \cline{8-8} 
$l_6$ & 20 & 0.5 & 100 & $-$ Step 4 $\rightarrow$ & 90 &  &  \\ \cline{1-4} \cline{6-6}
\end{tabular}
}
\centering
\label{tab:links}
\end{table}

\begin{table}[!htb]
\caption{Service attributes of the VNE example}
\begin{tabular}{|c|c|c|c|c|}
\hline
\textbf{Service} & \textbf{CPU} & \textbf{GPU} & \textbf{MEM} & \textbf{Allocated Node} \\ \hline
$s_1$ & 50 & 20 & 30 & $n_1$ \\ \hline
$s_2$ & 10 & 30 & 20 & $n_4$ \\ \hline
$s_3$ & 10 & 0 & 50 & $n_5$ \\ \hline
\end{tabular}
\centering
\label{tab:services}
\end{table}

\begin{table}[!htb]
\caption{Channel attributes of the VNE example}
\scalebox{0.85}{
\begin{tabular}{|c|c|c|c|c|c|}
\hline
\textbf{Channel} & $\mathbf{d^v_c}$ & $\mathbf{\rho^v_c}$ & $\mathbf{b^v_c}$ & \begin{tabular}[c]{@{}c@{}}\textbf{Pair}\\ \textbf{Quality-Revenue}\end{tabular} & \textbf{Path} \\ \hline
$c_1$ & 20 & 0.6 & 50 & 225 & $\{l_1 \rightarrow l_3, l_2 \rightarrow l_4\} $ \\ \hline
$c_2$ & 50 & 0.8 & 30 & 198 & $\{l_2 \rightarrow l_5\}$ \\ \hline
$c_3$ & 30 & 0.8 & 10 & 143.33 & $\{l_4 \rightarrow l_5, l_6\}$ \\ \hline
\end{tabular}
}
\centering
\label{tab:channels}
\end{table}

Before starting the allocation process, we sort the channel pairs (source and destination services together with the connection channel) according to their Quality-Revenue, obtaining an ordered list to  execute the anypath mapping, which for this case implies firstly $c_1 = (s_1, s_2)$, secondly $c_2 =(s_1, s_3)$ and thirdly $c_3 = (s_2, s_3)$ because $R^q_{c_1} > R^q_{c_2} > R^q_{c_3}$. In order to simplify the example calculations and for ease of understanding the algorithmic strategy, the revenue weight coefficients $\alpha_{x}$ and $\beta$ have been tuned equal to 1, while the quality-revenue parameter $\gamma$ has been configured to 500 to give considerable importance to the communication quality term compared to the previous parameters.

Therefore, we start in step 1 with channel $c_1$, whose destination service $s_2$ is not yet allocated. Thus, we look for the suitable nodes that meet the resource requirements, which in this case are $n_4$ and $n_5$. Hence, $s_2$ is associated to $n_4$ because it has the highest local PDR among the suitable nodes. Then, step 2 computes the anypath algorithm from suitable nodes $n_1$ and $n_5$ for $s_1$ to destination $n_4$ is computed, and $n_1$ with the virtual path $\{l_1,l_2,l_3,l_4\}$ is chosen because its anypath route is the only one that fulfills the maximum channel cost.

We follow with $c_2$, whose source service $ s_1$ has already been assigned. Therefore the anypath route is conversely calculated from the suitable destinations $\{n_2, n_5\}$ to the source $n_1$, taking into account that link $l_1$ is not available anymore because there is no more bandwidth free. Thereby, the established channel goes to $n_5$  across $\{l_2, l_5\}$ because it uses the lower number of physical links.

Finally, both ends $s_3$ and $s_2$ of $c_3$ are already allocated, thus we only need to run the anypath algorithm to find the shortest virtual path from $n_5$ to $n_4$ and verify that the attained cost is lower than the required. It is satisfactorily achieved using $\{l_4, l_5, l_6\}$ and the communication channel is enabled.

All this procedure  leads to a successful VN embedding that allows to run the VN application request using the topology mapping described in Tables \ref{tab:services} and \ref{tab:channels}, and updates the available resources of the substrate network in each iteration as shown in Tables \ref{tab:nodes} and \ref{tab:links}.

\section{Evaluation}\label{evaluation}

In this section, a simulation benchmark developed in MATLAB is presented to evaluate how the algorithm behaves in a realistic situation that allows to discuss the appropriateness for the target use case. A detailed profile of the setting is first described and then the results achieved are presented.

\subsection{Simulation Scenario}

\renewcommand{\thefootnote}{\fnsymbol{footnote}}
\footnotetext[1]{Random distributions with braces (\{\}) denote the integer domain.}

\begin{figure}[!ht]
    \centering
    \includegraphics[width=0.6\linewidth]{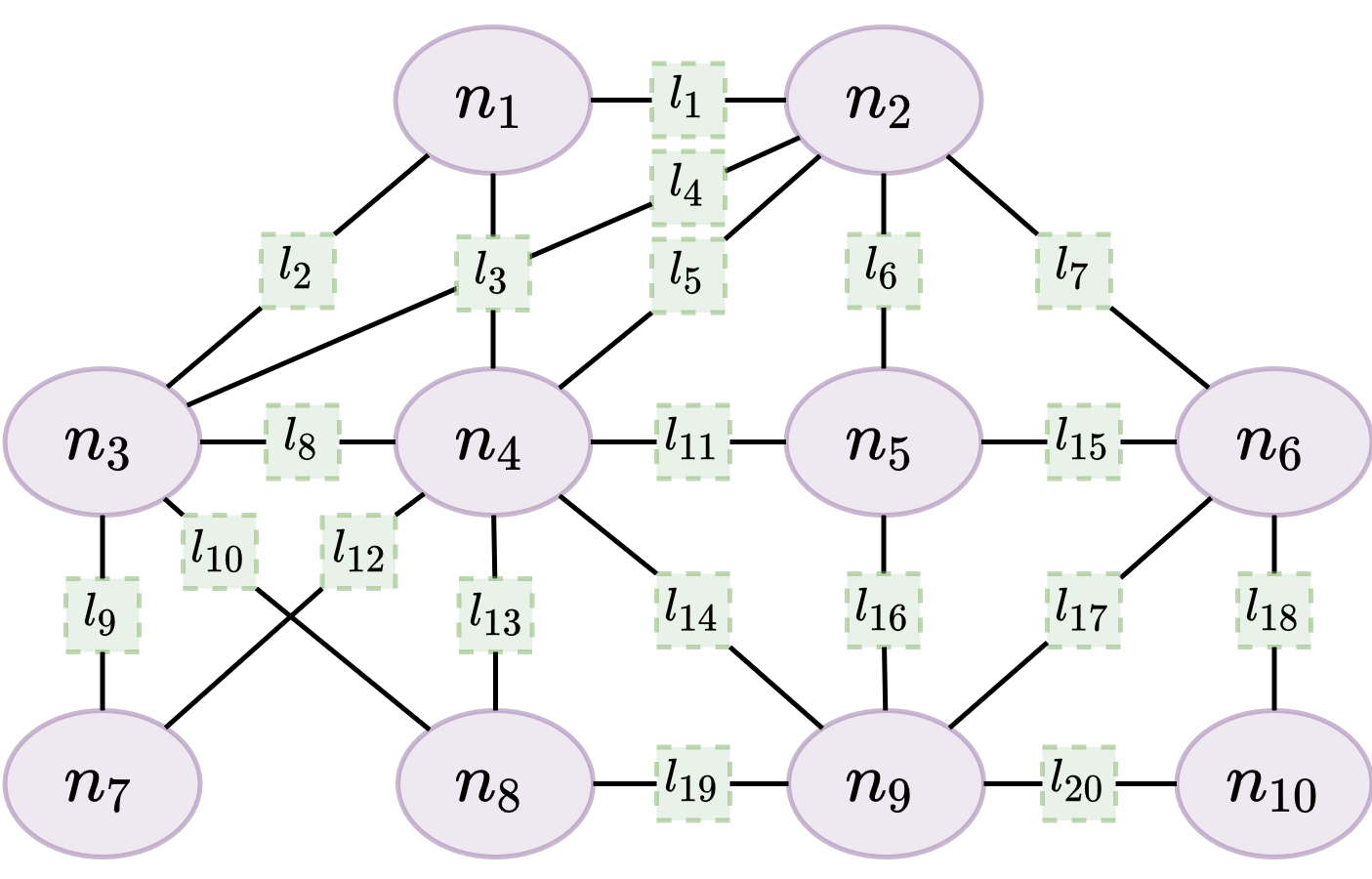}
    \caption{Diagram of the simulation substrate network.}
    \label{fig:simulation_substrate_network}
\end{figure}

We have defined a simulation environment for a generic substrate network depicted in Figure \ref{fig:simulation_substrate_network} and Tables \ref{tab:simulation_substrate_network_nodes} and \ref{tab:simulation_substrate_network_links}. The parameters of the node and link resources have been randomly chosen, thus having a substrate network that emulates a mesh swarm of drones with different amount of computing --- CPU, GPU and MEM --- and communication --- $b_l$ --- resources, and qualities of the communication links --- $d_l, \rho_l$.

\begin{table}[!htb]
\caption{Nodes of the simulation substrate network}
\begin{tabular}{|c|c|c|c|}
\hline
\multirow{2}{*}{\textbf{Node}} & \textbf{CPU} & \textbf{GPU} & \textbf{MEM} \\ \cline{2-4} 
 & $\sim U\{50, 150\}$\footnotemark{} & $\sim U\{30, 50\}$\footnotemark[\value{footnote}] & $\sim U\{50, 100\}$\footnotemark[\value{footnote}] \\ \hline
$n_1$ & 71 & 30 & 89 \\ \hline
$n_2$ & 98 & 41 & 85 \\ \hline
$n_3$ & 92 & 47 & 69 \\ \hline
$n_4$ & 136 & 45 & 81 \\ \hline
$n_5$ & 67 & 33 & 71 \\ \hline
$n_6$ & 84 & 30 & 79 \\ \hline
$n_7$ & 77 & 46 & 85 \\ \hline
$n_8$ & 119 & 50 & 55 \\ \hline
$n_9$ & 72 & 44 & 97 \\ \hline
$n_{10}$ & 132 & 36 & 100 \\ \hline
\end{tabular}
\centering
\label{tab:simulation_substrate_network_nodes}
\end{table}

\begin{table}[!htb]
\caption{Links of the simulation substrate network}
\begin{tabular}{|c|c|c|c|c|}
\hline
\multirow{2}{*}{\textbf{Link}} & \multirow{2}{*}{\textbf{Edges}} & $\boldsymbol{b_l}$ & $\boldsymbol{d_l}$ & $\boldsymbol{\rho_l}$ \\ \cline{3-5} 
 &  & $\sim U\{50, 100\}$\footnotemark[\value{footnote}] & $\sim U\{1, 10\}$\footnotemark[\value{footnote}] & $\sim U(0.9, 0.99)$ \\ \hline
$l_1$ & $(n_1, n_2)$ & 84 & 2 & 0.93 \\ \hline
$l_2$ & $(n_1, n_3)$ & 90 & 8 & 0.99 \\ \hline
$l_3$ & $(n_1, n_4)$ & 51 & 7 & 0.99 \\ \hline
$l_4$ & $(n_2, n_3)$ & 59 & 5 & 0.90 \\ \hline
$l_5$ & $(n_2, n_4)$ & 94 & 10 & 0.96 \\ \hline
$l_6$ & $(n_2, n_5)$ & 87 & 8 & 0.91 \\ \hline
$l_7$ & $(n_2, n_6)$ & 75 & 3 & 0.95 \\ \hline
$l_8$ & $(n_3, n_4)$ & 56 & 2 & 0.92 \\ \hline
$l_9$ & $(n_3, n_7)$ & 74 & 6 & 0.91 \\ \hline
$l_{10}$ & $(n_3, n_8)$ & 76 & 4 & 0.95 \\ \hline
$l_{11}$ & $(n_4, n_5)$ & 65 & 10 & 0.95 \\ \hline
$l_{12}$ & $(n_4, n_7)$ & 52 & 1 & 0.94 \\ \hline
$l_{13}$ & $(n_4, n_8)$ & 72 & 4 & 0.95 \\ \hline
$l_{14}$ & $(n_4, n_9)$ & 54 & 3 & 0.93 \\ \hline
$l_{15}$ & $(n_5, n_6)$ & 52 & 8 & 0.98 \\ \hline
$l_{16}$ & $(n_5, n_9)$ & 84 & 9 & 0.92 \\ \hline
$l_{17}$ & $(n_6, n_9)$ & 93 & 8 & 0.98 \\ \hline
$l_{18}$ & $(n_6, n_{10})$ & 56 & 2 & 0.96 \\ \hline
$l_{19}$ & $(n_8, n_9)$ & 56 & 9 & 0.96 \\ \hline
$l_{20}$ & $(n_9, n_{10})$ & 74 & 1 & 0.90 \\ \hline
\end{tabular}
\centering
\label{tab:simulation_substrate_network_links}
\end{table}

The performed simulation covers a single time window in which several VN requests are loaded to be handled by the embedding QRPAD-VNE algorithm. Specifically, the VN requests are generated randomly following the next directives, summarized in Table \ref{tab:simulation_virtual_network_configuration}:

\begin{itemize}
    \item Each VN request contains from 2 to 7 services uniformly distributed.
    \item Each service requires between 1 to 10 integer units of CPU and GPU, and between 1 to 5 integer units of MEM uniformly distributed .
    \item GPU resources are only requested by a 25\% of generated services (Bernoulli distribution).
    \item A communication channel between two services is established with a probability of 0.3 following the corresponding Bernoulli distribution.
    \item Each channel requires between 1 to 10 integer units of bandwidth $b^v_c$, a maximum communication delay $d^v_c$ of 10 to 50 integer units, and a reliability $\rho_c$ with a probability between 0.5 and 1. All the values are randomly chosen from their respective uniform distribution.
\end{itemize}

\begin{table}[!htb]
\caption{Configurations for a random VN request}
\begin{tabular}{|c|c|}
\hline
\textbf{Parameter} & \textbf{Configuration} \\ \hline
Number Services             & $\sim U\{2,7\}$\footnotemark[\value{footnote}]                \\ \hline
CPU                         & $\sim U\{1, 10\}$\footnotemark[\value{footnote}]              \\ \hline
GPU                         & $\sim U\{1, 10\}  \wedge B(0.25)$\footnotemark[\value{footnote}] \\ \hline
MEM                         & $\sim U\{1, 5\}$\footnotemark[\value{footnote}]              \\ \hline
Channel Connection          & $\sim B(0.3)$                   \\ \hline
$b^v_c$                       & $\sim U\{1, 10\}$\footnotemark[\value{footnote}]             \\ \hline
$d^v_c$                       & $\sim U\{10, 50\}$\footnotemark[\value{footnote}]            \\ \hline
$\rho^v_c$                       & $\sim U(0.5, 1)$                \\ \hline
\end{tabular}
\centering
\label{tab:simulation_virtual_network_configuration}
\end{table}

The functionalities associated with sensor and actuators, $f$ in the substrate network and $f^v$ in the virtual network, have been omitted to facilitate the interpretation of the results. Indeed, they would only introduce limitations in node assignment and do not influence algorithm performance itself.

Finally, weight coefficients of the Revenue $\alpha_{x}$ and $\beta$ have been set to 1 and 3 respectively, with the goal of increasing bandwidth relevance compared to the node resources. The equivalent coefficients of the Cost $\alpha_{x}'$ and $\beta'$ have also been set to 1 and 3 respectively since having the same values as in the revenue allows to study the normalized efficiency of the QRPAD-VNE algorithm. Otherwise, the $\gamma$ parameter of the
Quality-Revenue has been tuned to 3000 because the lowest value of the communication quality factor of a single channel $\frac{\rho^v_c}{d^v_c}$ is 0.01 and the corresponding highest value of the bandwidth $\beta \cdot b^v_c$ is 30, then a comparable relevance is given.

\subsection{Simulation Results}

\begin{figure}[!ht]
    \centering
     \begin{subfigure}{0.6\linewidth}
        \centering
        \includegraphics[width=\linewidth]{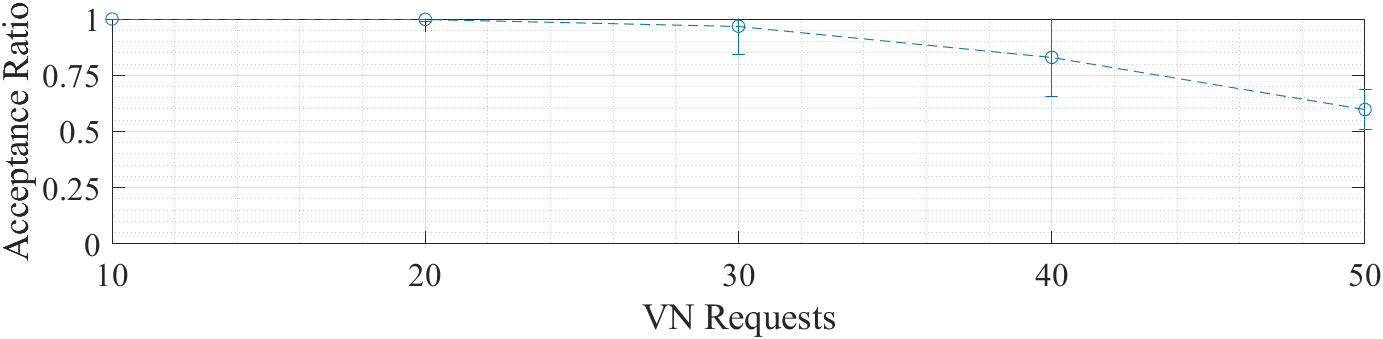}
        \caption{Average acceptance ratio.}
        \label{fig:acceptance_ratio}
     \end{subfigure}
     \hfill
     \begin{subfigure}{\linewidth}
        \centering
        \includegraphics[width=0.6\linewidth]{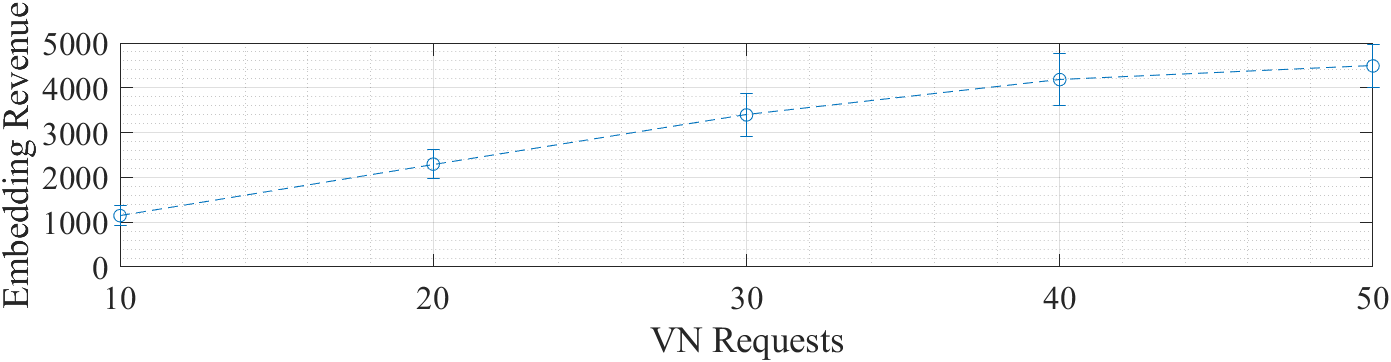}
        \caption{Average embedding revenue.}
        \label{fig:embedding_revenue}
     \end{subfigure}
     \begin{subfigure}{\linewidth}
        \centering
        \includegraphics[width=0.6\linewidth]{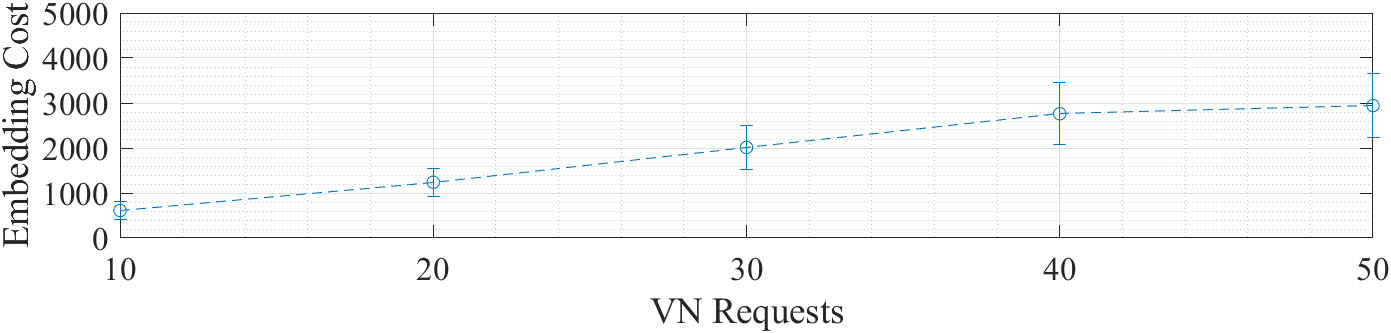}
        \caption{Average embedding cost.}
        \label{fig:embedding_cost}
     \end{subfigure}
     \begin{subfigure}{\linewidth}
        \centering
        \includegraphics[width=0.6\linewidth]{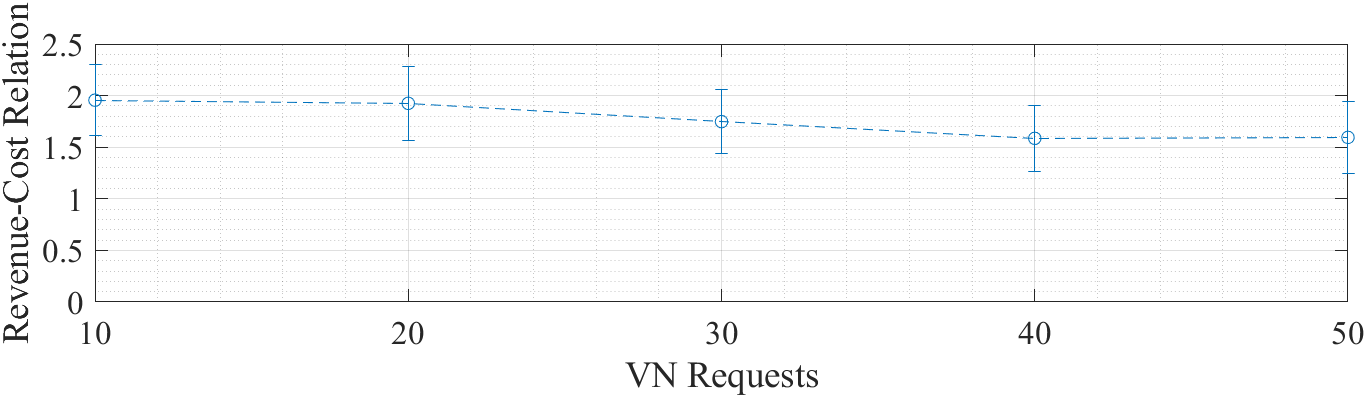}
        \caption{Average revenue-cost relation.}
        \label{fig:revenue_cost_relation}
     \end{subfigure}
    \caption{Metric results for the simulation.}
    \label{fig:metrics}
\end{figure}

\begin{figure*}[!ht]

    \begin{subfigure}{\linewidth}
       \centering
       \includegraphics[width=.329\linewidth]{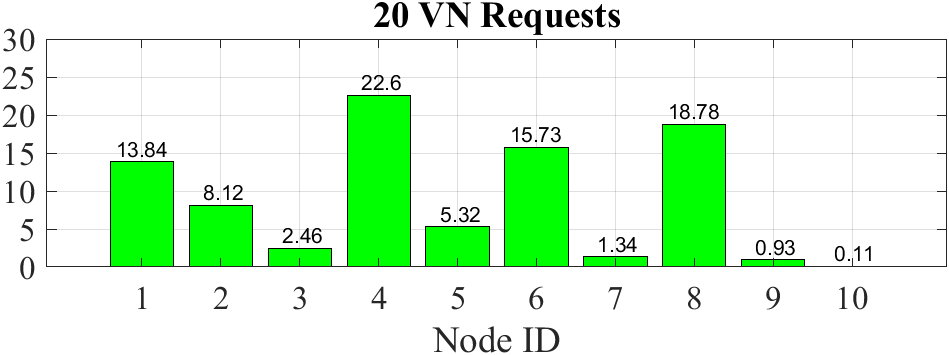}
       \includegraphics[width=.329\linewidth]{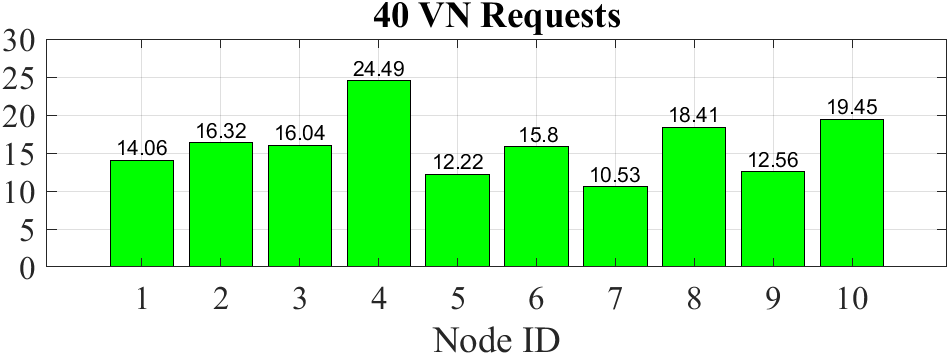}\\\vfill
       \includegraphics[width=.329\linewidth]{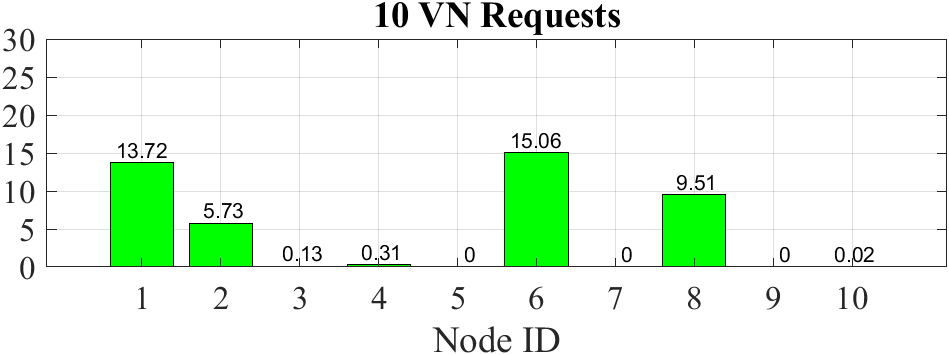}
       \includegraphics[width=.329\linewidth]{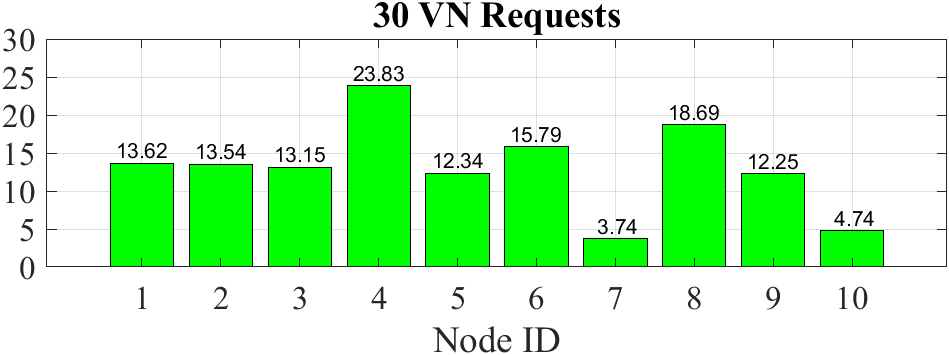}
       \includegraphics[width=.329\linewidth]{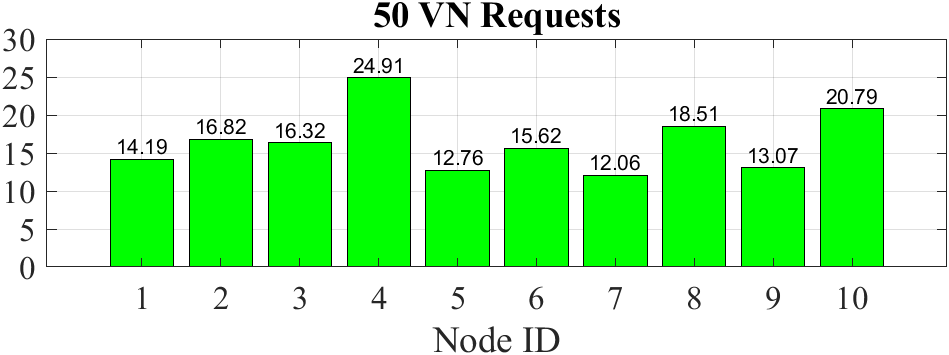}
       \caption{Average number of services allocated to each node.}
       \label{fig:service_node_usage}
    \end{subfigure}
    
    \begin{subfigure}{\linewidth}
       \centering
       \includegraphics[width=.329\linewidth]{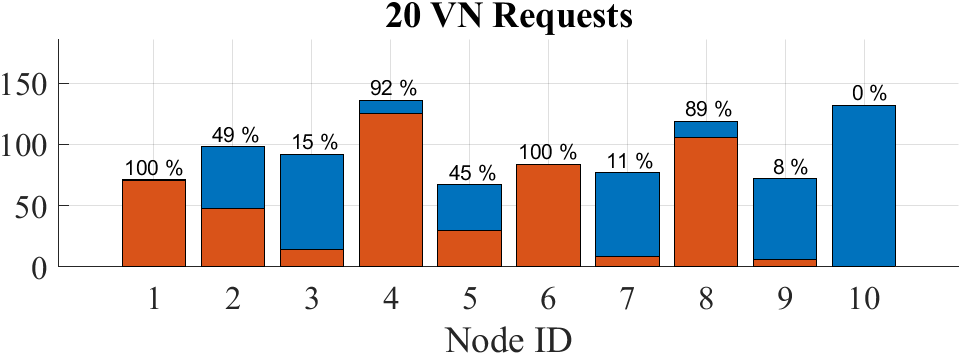}
       \includegraphics[width=.329\linewidth]{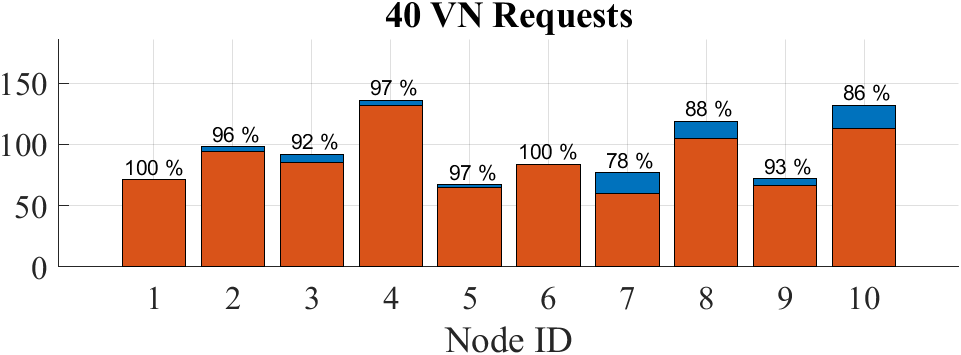}\\\vfill
       \includegraphics[width=.329\linewidth]{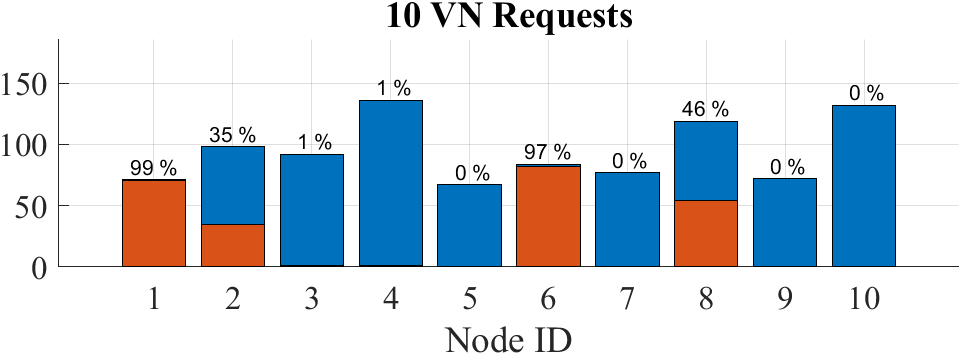}
       \includegraphics[width=.329\linewidth]{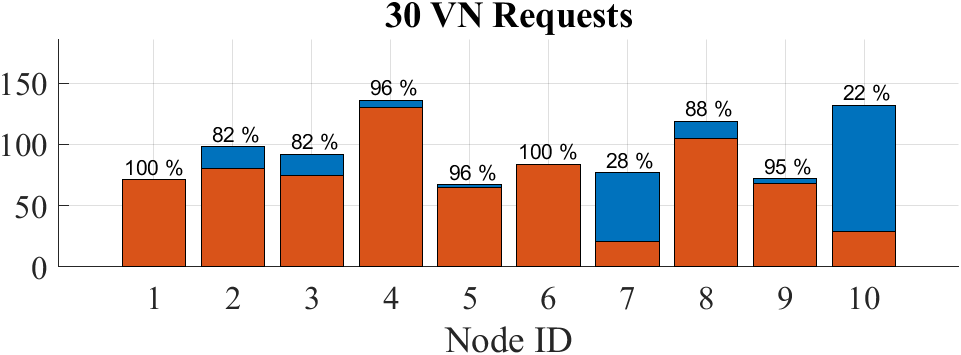}
       \includegraphics[width=.329\linewidth]{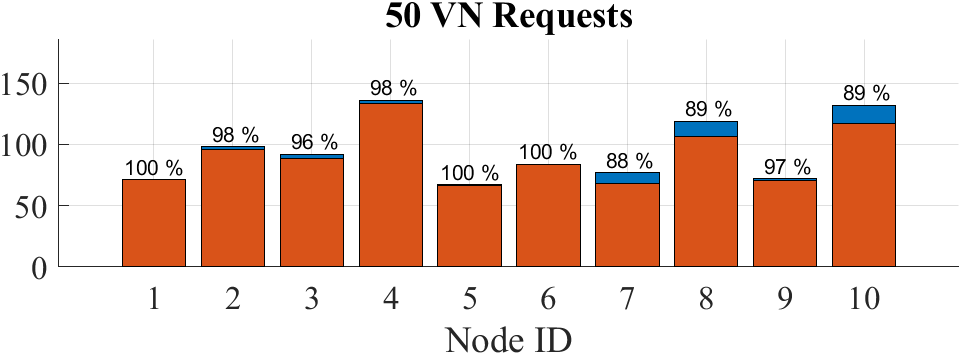}
       \caption{Average usage of the CPU of each node.}
       \label{fig:cpu_node_usage}
    \end{subfigure}
    
    \begin{subfigure}{\linewidth}
       \centering
       \includegraphics[width=.329\linewidth]{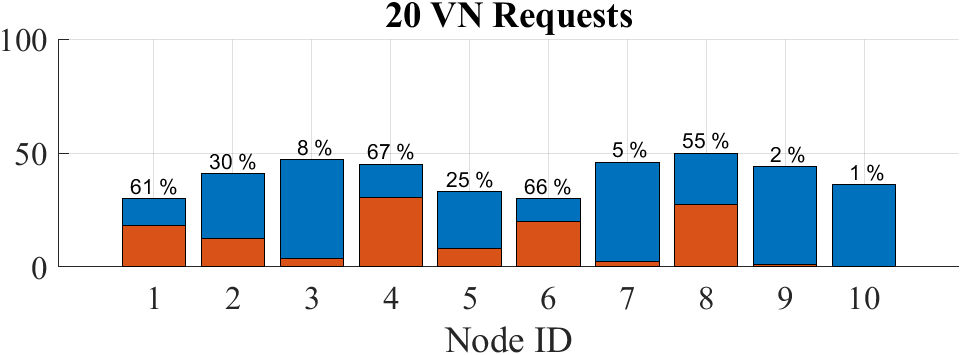}
       \includegraphics[width=.329\linewidth]{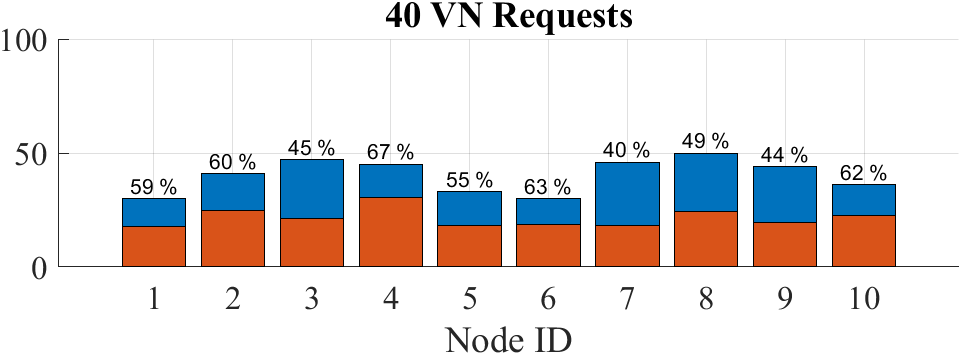}\\\vfill
       \includegraphics[width=.329\linewidth]{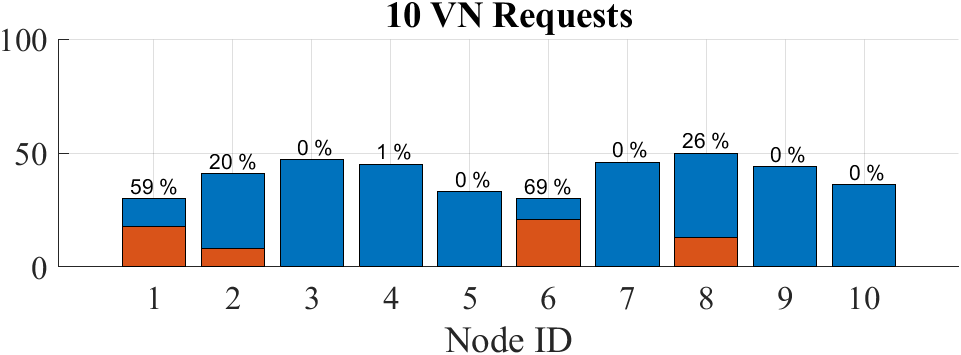}
       \includegraphics[width=.329\linewidth]{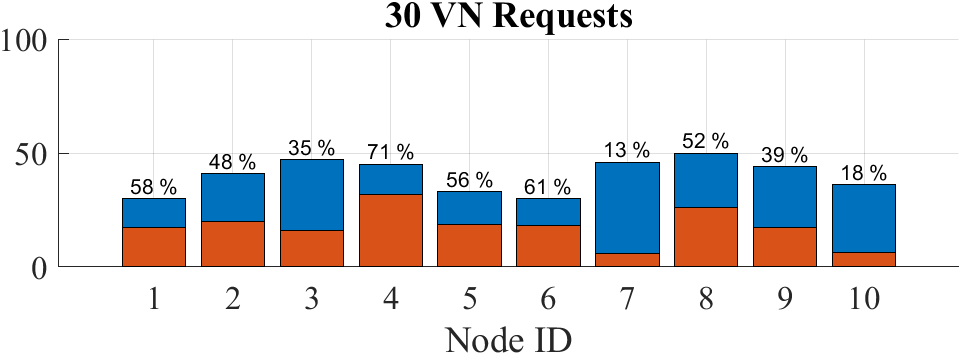}
       \includegraphics[width=.329\linewidth]{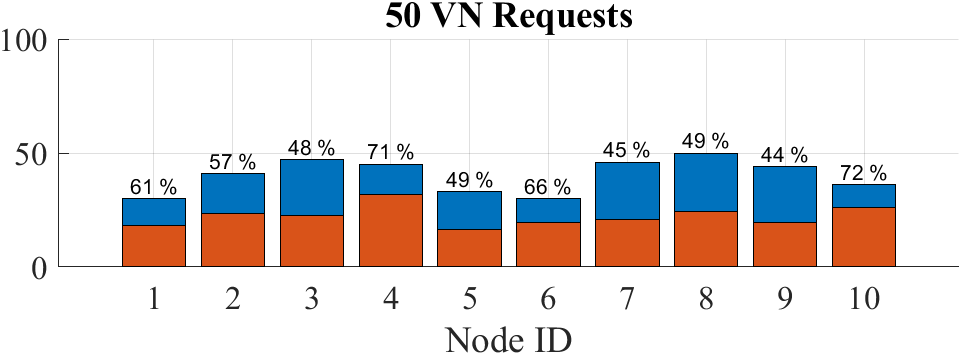}
       \caption{Average usage of the GPU of each node.}
       \label{fig:gpu_node_usage}
    \end{subfigure}
    
    \begin{subfigure}{\linewidth}
       \centering
       \includegraphics[width=.329\linewidth]{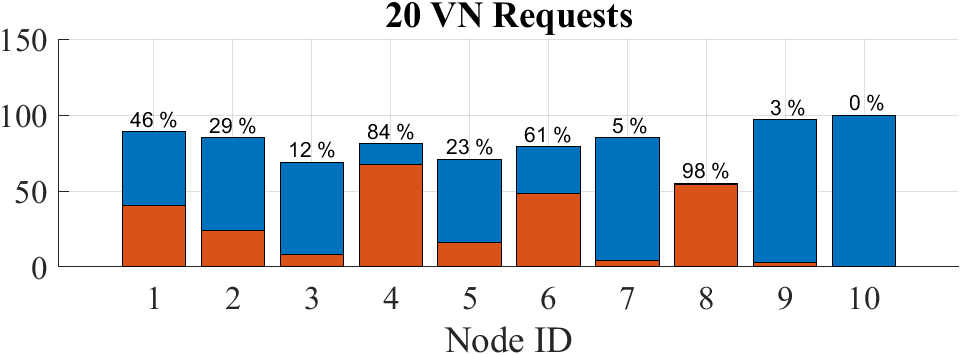}
       \includegraphics[width=.329\linewidth]{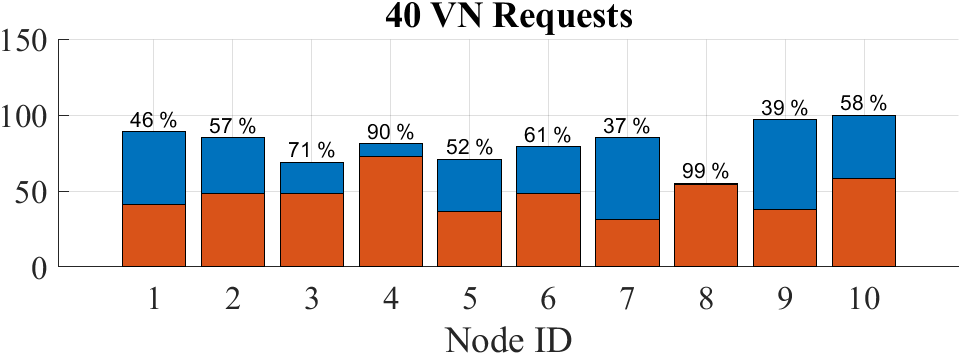}\\\vfill
       \includegraphics[width=.329\linewidth]{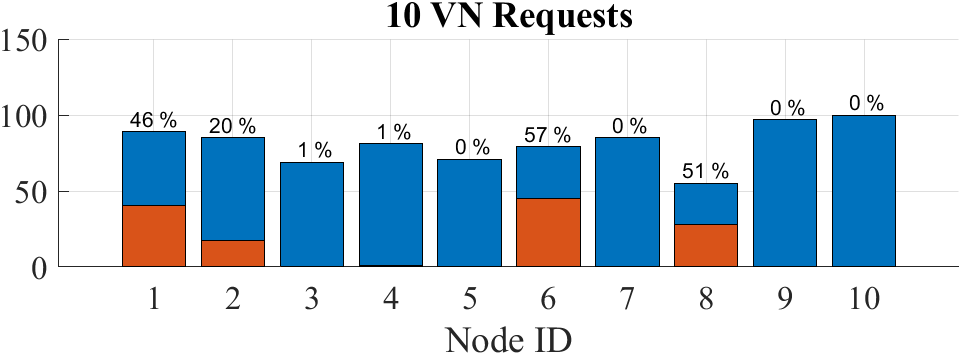}
       \includegraphics[width=.329\linewidth]{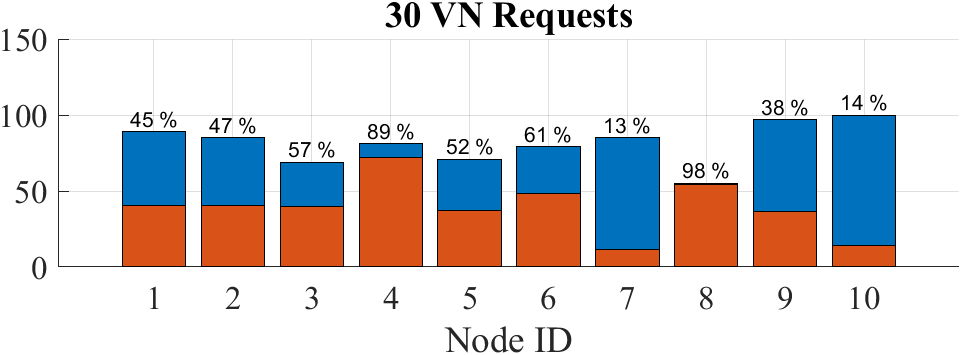}
       \includegraphics[width=.329\linewidth]{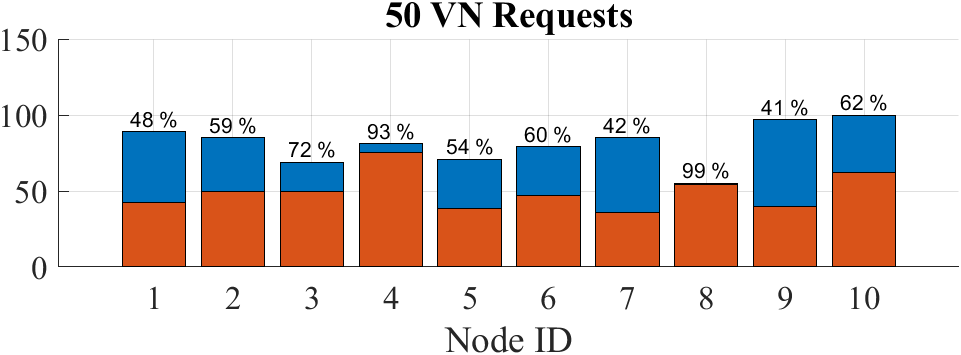}
       \caption{Average usage of the memory of each node.}
       \label{fig:mem_node_usage}
    \end{subfigure}
    
    \centering
    \caption{Average node usage metrics.}
    \label{fig:node_usage}
\end{figure*}

\begin{figure*}[!ht]
    
    \begin{subfigure}{\linewidth}
       \centering
       \includegraphics[width=.329\linewidth]{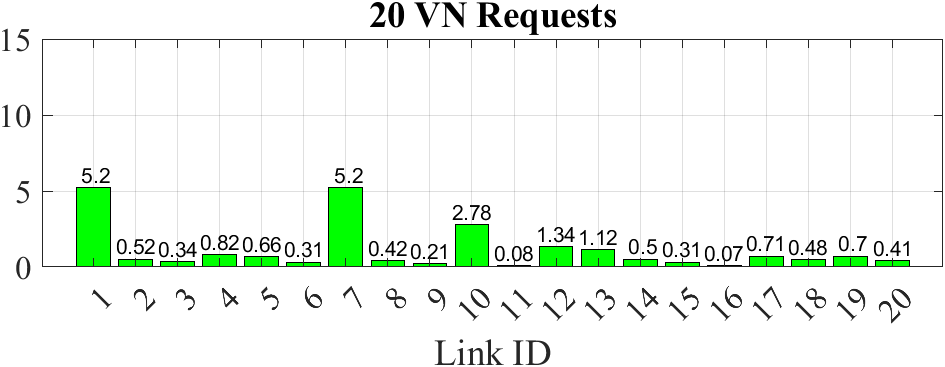}
       \includegraphics[width=.329\linewidth]{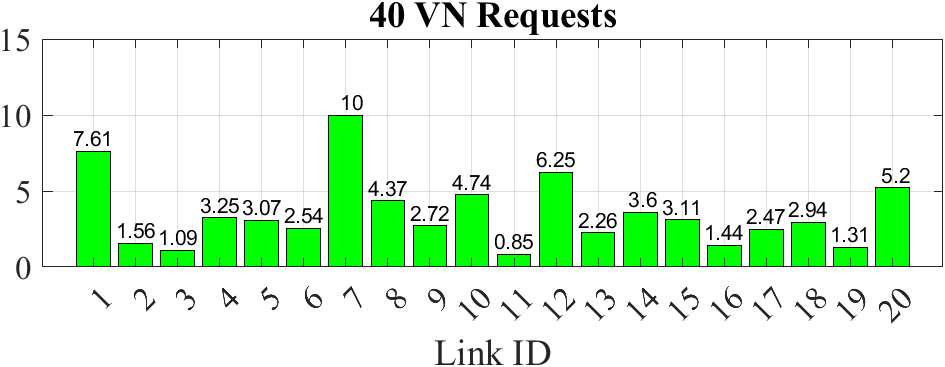}\\\vfill
       \includegraphics[width=.329\linewidth]{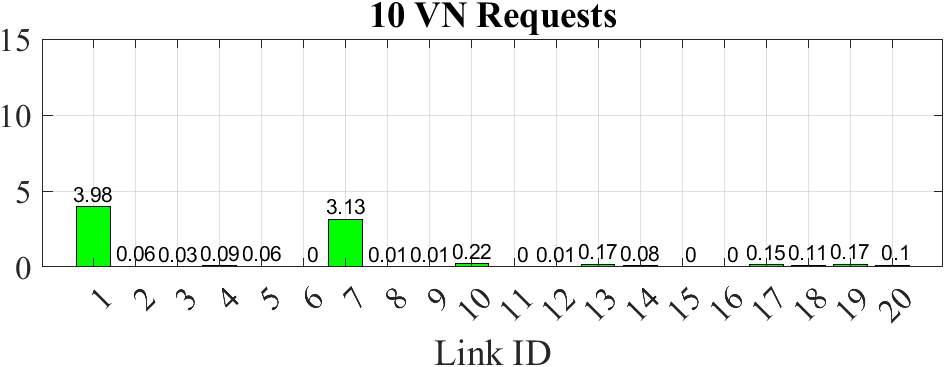}
       \includegraphics[width=.329\linewidth]{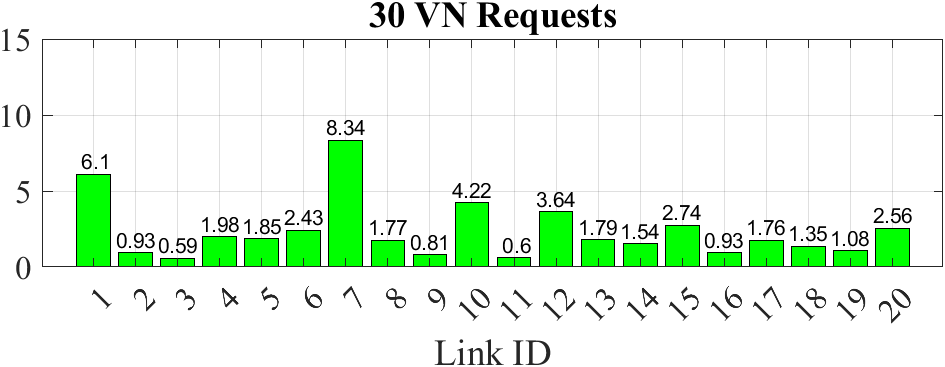}
       \includegraphics[width=.329\linewidth]{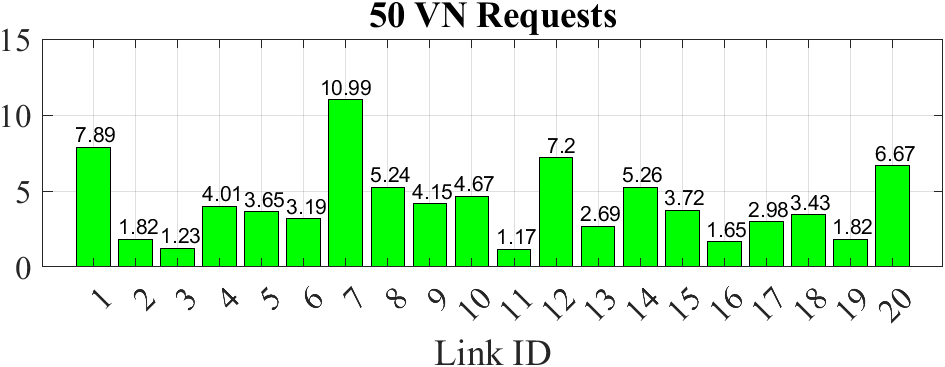}
       \caption{Average number of channels allocated to each link.}
       \label{fig:channel_link_usage}
    \end{subfigure}
    
    \begin{subfigure}{\linewidth}
       \centering
       \includegraphics[width=.329\linewidth]{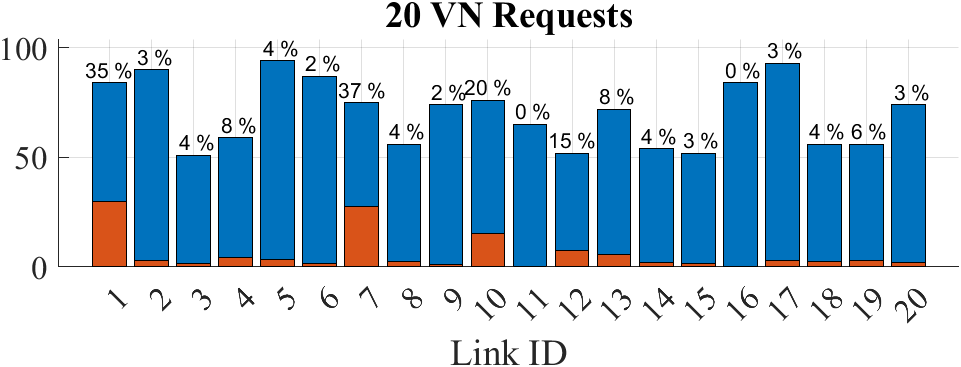}
       \includegraphics[width=.329\linewidth]{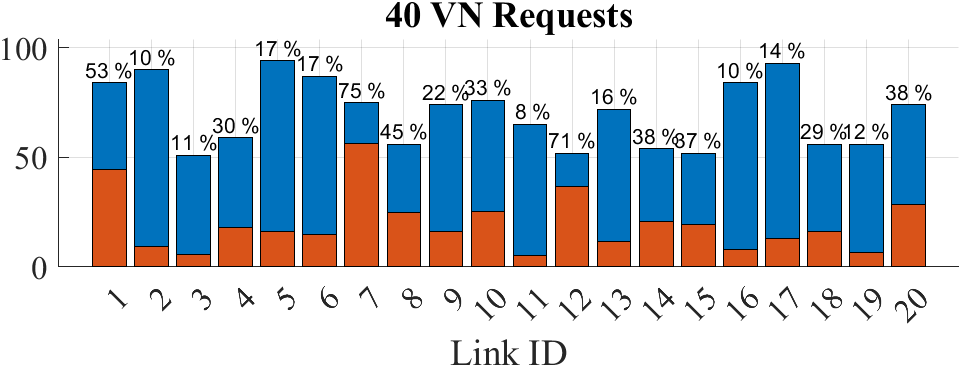}\\\vfill
       \includegraphics[width=.329\linewidth]{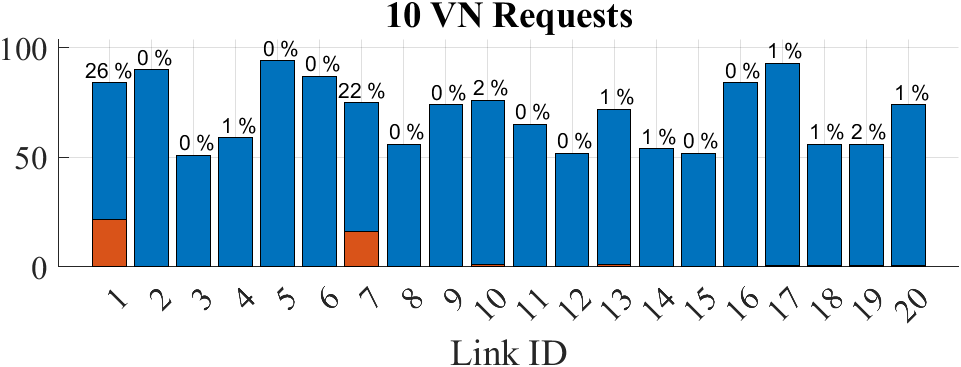}
       \includegraphics[width=.329\linewidth]{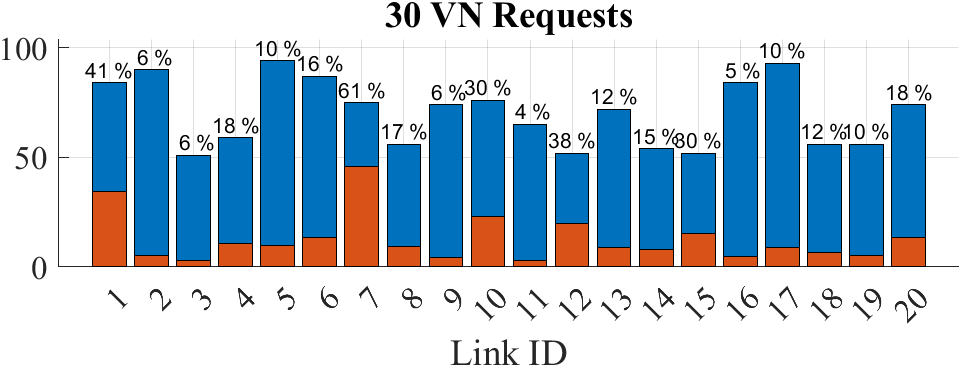}
       \includegraphics[width=.329\linewidth]{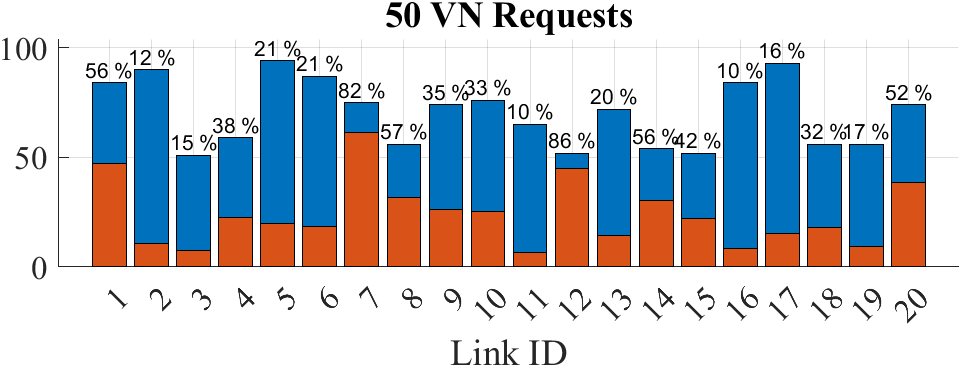}
       \caption{Average usage of the bandwidth of each link.}
       \label{fig:bw_link_usage}
    \end{subfigure}
    
    \centering
    \caption{Average link usage metrics.}
    \label{fig:link_usage}
\end{figure*}

According to the above simulation framework, we have completed several executions with different numbers of VN requests in order to study the behaviour of the algorithm according to the application load in a time window. Namely, the cases with 10, 20, 30, 40, and 50 have been studied, then 50 requests are randomly generated, but only the first specified number of them are used in the corresponding scenario. The attained results have been averaged over 100 independent iterations, and hereunder, the average performance of the algorithm for each aforementioned metric is calculated and the results are shown in Figure \ref{fig:metrics}. In addition, Figures \ref{fig:node_usage} and \ref{fig:link_usage} show the utilization of the nodes, links and their resources, after the allocation of each scenario, \deleted{for only three out of the five situations,} low load (10 VN requests), medium load (\added{20 and }30 VN requests) and high load (\added{40 and }50 VN requests) of the system.

Figure \ref{fig:acceptance_ratio} shows the average value of the Acceptance Ratio achieved in each scenario depending on the number of VN requests that feed the system, together with the standard deviation margin, to evaluate the ability to run multiple dataflow applications. In addition, Figures \ref{fig:embedding_revenue} and \ref{fig:embedding_cost} indicate the average Embedding Revenue and Cost, respectively, to be aware of the potential demand of the resources and the actual one, and later in Figure \ref{fig:revenue_cost_relation} the corresponding Revenue-Cost Relation is displayed to evaluate the capacity to use resources efficiently.

Regarding the usage of physical resources, Figure \ref{fig:service_node_usage} displays the average number of services that have been assigned to each substrate node in the three scenarios referred as relevant to evaluate the effect of the system load. Similarly, Figures \ref{fig:cpu_node_usage}, \ref{fig:gpu_node_usage} and \ref{fig:mem_node_usage} show the average percentage of resources of type CPU, GPU and memory, respectively, that is needed to allocate the number of chosen VN requests, as well as the average number of units that are used (red) over the total ones (blue).

Likewise, Figure \ref{fig:channel_link_usage} shows the average number of paths that each link in the substrate network has allocated and Figure \ref{fig:bw_link_usage} displays the average percentage of bandwidth resources that have been dedicated for the allocation of the communication channels, together with the average number of bandwidth units that are used (red) over the full capacity of the links (blue).

In view of the information given by the data charts, we can state some specific aspects about the performance of QRPAD-VNE algorithm for each metric relying on the state of use of the substrate network.

It can be seen that the Acceptance Ratio stands at around 1, with small variations, for situations with available resources (10, 20 and 30 VN requests), but it drops significantly for the 40-VN and 50-VN situation because the CPU resources start to scarce in the 30-VN scenario and it gets worse later, so it is difficult to find a combination of all resources for a very demanding request, which remains a resource bottleneck. Even so, smaller requests can find an allocation mapping successfully with a ratio ranging confidently above 50\% in high saturation situations.

As expected, Revenue and Cost increase with the number of VN requests in the system due to the higher demand of resources, but the Cost grows more slowly because some connected services are allocated in the same physical node and no network resource is needed for the corresponding channels. This effect is reflected in the Revenue-Cost Relation, which is always above 1. Nevertheless, it decreases slightly for heavy-loaded situations (50 and 40 requests) because some nodes and links of the substrate network are not suitable for the allocation anymore and farthest nodes are used and virtual paths are composed of more links. Although, it is not grossly affected by the multi-hop virtual paths because they are generally short.

Furthermore, as the number of VN requests grows, the resource utilization is distributed among all nodes and links in the substrate network, however some nodes, such as $n_1$, $n_2$, $n_6$ and $n_8$ are evidently more used than the others in low-load scenarios because they are preferred from the node ranking strategy based on local PDR. It is for this reason, resources of some links from these nodes ($l_1$ and $l_7$ for 10 VN requests), for the most part, are most commonly consumed.

The present proposal provides a simulation framework based on number of VN requests in the time window to evaluate, while compared to previous works in the field of wireless VNE, such as \cite{application_vne_industrial_wireless} and \cite{reliable_vne_atn}, they apply a stochastic arrival mechanism over time, and the parameters used in the evaluation of the metrics are selected differently. Thus, quantitative comparison is not possible, but qualitatively we can assert that our algorithm takes into consideration the demanding revenue of each request to optimize allocation and avoid fragmentation, achieving a very successful Acceptance Ratio. Moreover, wireless links are mathematically modeled in a way physical wireless effects, such as interference, are simplified to numbered quality of the links.

\section{Conclusion}\label{conclusion}

This paper proposes the usage of VNE strategies as one of the first steps to distribute applications among a set of nodes (UAVs) in a FANET, allowing the implementation of complex applications that can not be executed in just one node, and that may not tolerate the latency or the required bandwidth to be implemented in a remote cloud node. 

The presented QRPAD-VNE algorithm assigns physical resources to the virtual requests, taking into account the requirements of the virtual requests in terms of functional (e.g. a particular sensor or a discrete feature of the node) and non-functional resources (e.g. the computing power or the available memory). Furthermore, it takes advantage of the broadcast nature of wireless communications to create robust communications by using the anypath routing mechanism to build redundant links. In addition, the algorithm uses a ``theoretical cost'' that is used to maximize theoretical revenues in order to decide how requests are assigned to resources. However, the mobility aspect of FANETs remains open, which now is contemplated through the recomputation of the VNE algorithm using the live applications to command a migration of services according to changes in the network topology if mobility occurs.

The proposal is formalized and validated through simulations, showing its adequacy for deploying complex distributed applications in a FANET scenario. We have observed how it can assign resources even when there is a large number of requests.

We must also highlight the importance of the work with regard to the implementation of a basis framework for the modeling and execution of VNE in wireless environments, because it supports the definition of multiple types of resources in the graph model and network communications are regarded as a graph routing problem. Besides, including the dataflow programming model as virtual requests must be emphasized since it provides a convenient and appropriate way to implement final applications in sensing IoT networks.

As future work, several significant research and development lines are open to improve the design of the wireless VNE framework presented. On the one side, the time variable must be incorporated to it, both in the lifecycle of the VN requests (dataflow applications) and in the simulation tool to get a dynamic time window that manages the arrival of new requests, thus we could evaluate the performance over time. On the other side, the intrinsic mobile nature of FANET nodes shall be considered in the VNE algorithm through the integration of a mobility model, which implies the deploymet of services according to potential mobility which does not disrupt the service availability. It is also important the adoption of a stochastic model in the resource management, either in the simultaneous usage of node resources and operational functionalities, and in the level of compliance to apply QoS policies in the communication links. Besides that, as the main focus of this work is the allocation of a communication path (channels into links), therefore, next steps should also consider the inclusion of a node ranking strategy to define a more advanced cost to prioritize among nodes in service allocation, for instance related to energy consumption, usage cost of the hardware resources or quality of data generated by required sensors.

\section*{Acknowledgment}

This research has been promoted by the Spanish Ministry of Science, Innovation and Universities, under the Formación de Profesorado Universitario (FPU, FPU19/01284) fellowship, and it has been supported by the DocTIC PhD program of the University of Vigo and the IACOBUS exchange program of the European Association for Galicia-North of Portugal Region Cooperation (GNP-AECT). It has received funding from the European Union’s Horizon 2020 research and innovation program ECSEL Joint Undertaking (JU) under grant agreement No. 876487, NextPerception project—“Next Generation Smart Perception Sensors and Distributed Intelligence for Proactive Human Monitoring in Health, Wellbeing, and Automotive Systems”, cofunded by the Spanish Ministry of Science and Innovation (PCI2020-112174), and by the Spanish Government under research projects “Enhancing Communication Protocols with Machine Learning while Protecting Sensitive Data (COMPROMISE)" (PID2020-113795RB-C33/AEI/10.13039/501100011033) and "AriSe1: Redes Ultradensas sin Celdas (DeCk)" (PID2020-116329GB-C21). It has also been supported by the European Regional Development Fund (ERDF) and the Galician Regional Government, under the agreement for funding the Atlanttic Research Center for Information and Communication Technologies (AtlantTIC) and the "Grupo de Referencia Competitiva (GRC)" (GRC-ED431C 2022/04 T254). In addition, it has been funded for open access by the Universidade de Vigo/CISUG agreement with Elsevier.


\bibliographystyle{elsarticle-num}

\bibliography{bibtex}

\end{document}